\documentclass[pra,preprint,
final,showpacs,superscriptaddress,nofootinbib]{revtex4}

\usepackage{graphics}
\usepackage{amsmath,amssymb}
\usepackage{graphicx}% Include figure files
\usepackage{dcolumn}% Align table columns on decimal point
\usepackage{bm}% bold math

\begin{document}

\title{Classical dynamics of a charged particle in a laser field beyond the dipole approximation}

\author{Paul Jameson}
\affiliation{School of Mathematics and Statistics,
             University of Plymouth,
             Plymouth, PL4 8AA,
             United Kingdom}
\author{Arsen Khvedelidze}
\affiliation{Department of Theoretical Physics,
             A.Razmadze Mathematical Institute,
             Tbilisi, GE-0193,
             Georgia}
\affiliation{Laboratory of Information Technologies,
             Joint Institute for Nuclear Research,
             Dubna, 141980,
             Russia}

\date{April 2008}

\begin{abstract}
The classical dynamics of a charged particle travelling in a laser
field modelled by an elliptically polarized monochromatic
electromagnetic plane wave is discussed within the time
reparametrization invariant form of the non-relativistic
Hamilton-Jacobi theory. The exact parametric representation for a
particle's orbit  in an arbitrary plane wave background beyond the
dipole approximation and including effect of the magnetic field is
derived. For an elliptically polarized monochromatic plane wave the
particle's trajectory, as an explicit function of  the laboratory
frame's time, is  given in terms of the Jacobian elliptic functions,
whose modulus is proportional to the laser's intensity and depends
on the polarization of radiation. It is shown that the system
exposes the \textit{intensity duality}, correspondence between the
motion in the backgrounds with various intensities. In virtue of
the modular properties of the Jacobian functions, by starting with
the representative ``fundamental solution '' and applying a
certain modular transformations one can obtain the particle's orbit in
the monochromatic plane wave background with arbitrarily
prescribed characteristics.
\end{abstract}

%\keywords{}


\maketitle \tableofcontents

%%%%%%%%%%%%%%%%%%%%%%%%%%%%%%%%%%%%%%%%%%%%%%%%%%%%%%%%%%
\section{Introduction}
%%%%%%%%%%%%%%%%%%%%%%%%%%%%%%%%%%%%%%%%%%%%%%%%%%%%%%%%%%

By the end of the 20th century Quantum Electrodynamics set an
undoubted example of a theory allowing a thorough understanding and the ability to make predictions to
an arbitrary precision the interaction of individual fundamental
particles, photons and electrons. Until the development of the laser existing photon sources
produced incoherent photons one at a
time and the perturbative QED, based on an elementary
\textit{single photon} interacting with a {\em single electron},
was a perfectly adequate and accurate theory despite not taking into
account the \textit{ coherence} aspect of the photons. Since the
1960's, after the advances in laser technology, the
first theoretical studies of collective multi-photon relativistic
dynamics began. Over the past fifteen years the revolutionary
invention for a new technique of laser pulse amplification has led to
the possibility to create the attosecond pulses of light beams of a
very high intensity and made the practical study of a new
area of QED, \textit{physics of high-intensity laser-matter
interactions} possible (see e.g., textbooks \cite{books} and  recent
surveys
\cite{SalaminHuHatsagortsyanKeitel},\cite{MourouTajimaBulanov}).
Contemporary renaissance of these research activities has opened up
many new promising opportunities to develop our  understanding of
fundamental physics as well as the possibility of absolutely new
applications.

The necessity to modify  the conventional single particle
description of radiation-matter interaction  at high intensities
is arguably best illustrated by classical Thomson scattering, i.e.
scattering  of an electromagnetic wave by a free electron. The
theory of ordinary Thomson scattering \cite{Heitler,Jackson}
assumes that a linearly polarized monochromatic plane wave's phase
\footnote{The monochromatic electromagnetic plane wave represents
the simplest way to mathematically model a laser field, which is a
reasonable assumption providing the transverse directions of the
laser beams are much larger than the dimensions of the system
considered.} is approximated as $\omega_Lt-
\boldsymbol{k}_L\cdot\boldsymbol{x} \approx \omega_Lt $ ({\em
dipole approximation}) and the particle at rest responds only to
the electric field
\[\boldsymbol{E}=\boldsymbol{E}_0\cos\omega_Lt\,.
\]
Therefore the electron consequently executes simple harmonic
motion at the same frequency as the wave, $\omega_L$, along the electric
field direction:
\begin{equation}\label{eq:traj}
\boldsymbol{x}=-
\frac{e}{m\omega_L^2}\,\boldsymbol{E}_0\,\cos\omega_Lt\,.
\end{equation}
The radiation from such an oscillating charge is calculated with  the
aid of the classical Larmor radiation formula \cite{Jackson} which
results in the universal, frequency independent,   total cross
section
\begin{equation}\label{eq:Thomson}
    \sigma_{\mathrm{T}}=\frac{8\pi}{3}\,
    \left(\frac{e^2}{mc^2}\right)^2\,.
\end{equation}
The assumption that the magnetic part of the Heaviside-Lorentz
force of the plane wave,
\[
\boldsymbol{F} = e\, \boldsymbol{E}+
\frac{e}{c}\,\boldsymbol{v}\,\times\,\boldsymbol{B}\,,
\]
can be ignored,
$$
 \frac{v}{c}\,||\boldsymbol{B}_0|| \ll  ||\boldsymbol{E}_0||\,,
$$
implies, taking into account the equality,
$||\boldsymbol{E}_0||=||\boldsymbol{B}_0||\,,$
  that the electron motion should be completely
non-relativistic,  $v \ll c\,$. According to (\ref{eq:traj}) the
maximum electron velocity is $v_{max}= e
{||\mathbf{E}_0||}/{m\omega_L}\, $ and therefore the
non-relativistic character of motion will hold true providing
\begin{equation}\label{eq:intensity}
\eta^2:= \frac{e^2}{\omega^2_L m^2c^2}\,\boldsymbol{E}_0^2=
\frac{2}{\pi}\,\frac{e^2\lambda^2_L}{m^2c^5}\,{I}_L \ll 1\,.
\end{equation}
Here $\lambda_L$ is the radiation wavelength and the beam intensity
 ${I}_L:={c}\,\boldsymbol{E}_0^2/{8\pi}$ has been introduced.

When the dimensionless \textit{intensity parameter} $\eta$ is
sufficiently large the dipole approximation is no longer valid and
magnetic force cannot justifiably be ignored. The electron motion
becomes a nonlinear function of the driving force and relativistic
effects will modify photon-electron scattering. The classical
fully relativistic four-dimensional treatment of the
light-electron scattering when the electromagnetic radiation is
modelled by an electromagnetic monochromatic plane is well
established. In this background the classical relativistic
equation of motion for the electron with the full
Heaviside-Lorentz force and taking into account the retardation
effect  can be solved exactly. This provides with an implicit
parametric representation for the electron trajectory\footnote{To
the best of our knowledge J.Frenkel in 1925 was the fist one who
presented the parametric relativistic solution \cite{Frenkel}.
Since that publication this problem has been studied many times in
different context (for the principal references see reviews
\cite{SalaminHuHatsagortsyanKeitel},\cite{MourouTajimaBulanov} and
original papers \cite{classsol}) and became at issue for classical
textbooks \cite{LandauLifshitz,ItzyksonZuber,Thirring}. The brief
history of strong-field physics written by the pioneer of this
area can be found in the recent book \cite{ReissHistory}. }. In
this solution the relativistic particle's position vector
$\boldsymbol{x}$ is given as a function of the proper time which
is in turn a function of $\boldsymbol{x}$. The explicit solution
$\boldsymbol{x}(t)$ is a subtle function of the physical time,
known only in the form of an infinite series expansion over the
harmonics with a fundamental frequency depending on the radiation
intensity \cite{Sengupta1949}. The  appearance of the higher
harmonics in the scattered radiation is a direct manifestation of
this fact and results in the modification of  the classical
Thomson cross section (\ref{eq:Thomson}) for a high-intensity
laser beam ( see, for example, the basic references
\cite{Sengupta1949,VachaspatiSarachikSchappert1970} and for the
recent direct experimental observation of the second harmonics
\cite{ObserHarmonics}).

In order to fully understand the relativistic effects it is essential to have a clear understanding of how the  transition from
non-relativistic motion to the relativistic motion occurs. Analysis
(see for details~\cite{Reiss,Reiss00}) of the deviations from the
non-relativistic regime shows the possibility to categorize the
relativistic corrections into two groups; the first one is due to
the influence of the magnetic part,
$\boldsymbol{v}\,\times\,\boldsymbol{B}\,,$ of the
Heaviside-Lorentz force and gives the first-order relativistic
effects $O(v/c)\,.$ The second category are the ``true relativistic''
second-order effects $O(v^2/c^2)$ such as, for example, the well-known
relativistic corrections to the kinetic energy, Darwin term and
spin-orbit coupling. It was shown by H.Reiss
\cite{Reiss,Reiss00}, that there is an intermediate regime with a
wide range of parameters (in terms of the laser's intensity $\&$ frequency) where
the leading contribution to the deviation from the non-relativistic
description comes from the magnetic fields effects, while the
"truly relativistic" $v^2/c^2$ effects can be still be neglected. This important
observation gives justifies exploiting the combined methods,
where the relativistic effects are taken into account only partially
 as corrections. Such an approach is, in particular, very
attractive to the description of laser-atom interactions, where
the whole formalism is intrinsically non-relativistic (see the
recent publication  \cite{keitel07} and references therein).
 It is also
important since the widely used numerical technique becomes very
cumbersome when passing to the fully four dimensional relativistic
covariant description.

Several partially relativistic approaches, taking into account
magnetic field's influence on the particle's dynamics were developed
(see, for example,\cite{Reiss,Reiss00,Salamin} and
references therein). The straightforward way to take into account
the magnetic field is to go beyond the dipole approximation by
including the full phase dependence $\omega_Lt-
\boldsymbol{k}_L\cdot\boldsymbol{x}$ in the corresponding plane
wave potential\footnote{Note however, the discussions in
\cite{nonrel1,nonrel2} of the possible artifacts arising from this
way of partial relativistic consideration as well as necessity to
include the radiation damping effects
\cite{Sanderson,SteigerWoods}. }. Since the exact parametric
solution to the relativistic Lorentz equation for a charged
particle in plane wave background is known one could naively expect
that the corresponding Newton equation can be solved exactly in
an even more simple form. However, surprisingly, we were not able to
find (in the extensive literature on laser-matter interactions) such
an exact solution to Newton's equations of motion for a charged
particle in plane wave background. In this note we aimed to fill
this gap by providing the description of the intermediate region of the
laser-particle interaction and to present an exact parametric
solution to the non-relativistic equations  of motion that is in a
close analogy  to the corresponding relativistic problem. As an
application of the general formulae we will consider in detail the case of an
elliptically polarized monochromatic plane wave. For this case we
derive the explicit representation of a charged particle's orbit
in terms of the laboratory frame's time. The solution is given in
terms of the Jacobian elliptic functions, whose modulus depends on
the background's radiation intensity and polarization. The particle's
motion represents  a drift displacement and an infinite sum of
harmonic oscillations with the fundamental oscillation frequency depending on the radiation intensity in a nonlinear way.
Furthermore, owing to the modular properties of the Jacobian
functions which is the existence of relations between elliptic
functions  with different periods, we attest an interesting
\textit{duality} between  motions in backgrounds with a various
intensities regimes. In particular, we show how a particle's
trajectory in a monochromatic plane wave with a high intensity can
be obtained from the \textit{``fundamental solution,''} which describes the motion in a background with low intensity.

The presentation of the material in the article is gathered as follows.
In Section \ref{sec2} the non-relativistic motion
of a charged particle in an external electromagnetic background is
reformulated in a time reparametrization invariant fashion. The equal
footing of time and space coordinates mimics the relativistic
theory and enables us to use the conventional stationary
Hamilton-Jacobi method to find a parametric solution to the
non-relativistic equations of motion for a charged particle
travelling in an arbitrary plane wave background. In Sections
\ref{sec3} and \ref{sec4} we discuss in details a particle
trajectories in the  monochromatic plane wave background with an
arbitrary elliptic polarization. In Appendix \ref{sec:appendix1}
we briefly comment on the instantaneous implementation of a gauge
symmetry and Galilean boost transformation for a
 classical ``non-relativistic''  particle in an
electromagnetic background. Appendix \ref{ap:genericbound}
sketches the derivation of a particle's trajectory for generic
boundary conditions. Finally, in order to make article self-contained, Appendix \ref{sec:appendix2} gives  some mathematical
features of the Jacobian elliptic functions used throughout the main text.

%%%%%%%%%%%%%%%%%%%%%%%%%%%%%%%%%%%%%%%%%%%%%%%%%%%%%%%%%%
\section{Non-relativistic particle in an external field }\label{sec2}
%%%%%%%%%%%%%%%%%%%%%%%%%%%%%%%%%%%%%%%%%%%%%%%%%%%%%%%%%%%%%%%%%%

A point non-relativistic particle, with mass $m$ and electric
charge $-e$, moving in an external electric field $\boldsymbol{E}$
and magnetic field $\boldsymbol{B}$ is influenced by the
Heaviside-Lorentz force. A particle's trajectory
$\boldsymbol{x}(t)$ may be determined from Newton's equations of
motion
\begin{equation}\label{eq:NE}
    m\, \frac{\mathrm{d}^2\boldsymbol{x}(t)}{\mathrm{d}t^2}\,=
    e\,\boldsymbol{E}(t, \boldsymbol{x}(t)) +
    \frac{e}{c}\,\frac{\mathrm{d}\boldsymbol{x}}{\mathrm{d}t}\,\times
    \boldsymbol{B}(t,\boldsymbol{x}(t))\,.
\end{equation}
The nonlinear equations (\ref{eq:NE}) can be reproduced within the
conventional variational principle  of least action based on the
following ``non-relativistic''  Lagrangian function\footnote{The
name ``non-relativistic'' is somewhat misleading because the
Lagrangian (\ref{eq:nrL}) is not Galilean invariant, one can speak
only about an approximate Galilean symmetry for small particle
velocities (see discussion in \cite{Rohrlich} and  Appendix
\ref{sec:appendix1}).}
\begin{equation}\label{eq:nrL}
    \mathcal{L}\left(\boldsymbol{x}\,,
    \frac{\mathrm{d}\boldsymbol{x}}{\mathrm{d}t}\,, t\right) = \frac{m}{2}\,
    \frac{\mathrm{d}\boldsymbol{x}}{\mathrm{d}t}\cdot\,\frac{\mathrm{d}\boldsymbol{x}}{\mathrm{d}t}+
    \frac{e}{c}\,\frac{\mathrm{d}\boldsymbol{x}}{\mathrm{d}t}\cdot\,\boldsymbol{A}(t, \boldsymbol{x}(t))
- e\,\Phi(t,\boldsymbol{x}(t)) \,
\end{equation}
 if the external electric field $\boldsymbol{E}$ and
magnetic field $\boldsymbol{B}$ are defined in terms of the gauge
potential $A^\mu(t, \boldsymbol{x})=(\Phi(t,\boldsymbol{x})\,
,\boldsymbol{A} (t,\boldsymbol{x}))$ in the standard way
\begin{eqnarray}\label{eq:EB}
    \boldsymbol{E}(t,\boldsymbol{x})&=& \nabla \Phi(t,\boldsymbol{x})-
    \frac{1}{c}\frac{\partial}{\partial t}\,\boldsymbol{A}(t,\boldsymbol{x})\,, \\
     \boldsymbol{B}(t,\boldsymbol{x})&=& \nabla\mathbf{\times}\, \boldsymbol{A}(t,\boldsymbol{x})
     \,.
\end{eqnarray}
Here we intend to solve (\ref{eq:NE}) for a special case of
idealized laser field described by the  so-called electromagnetic
{\it monochromatic plane wave}. However, before restricting ourselves to
this special case, we consider at first the more general plane
wave background  \cite{LandauLifshitz} with a gauge potential of the
form
\begin{equation}\label{eq:plwave}
A_\mu(\boldsymbol{x}, t)=A_\mu(\xi)\,,
\end{equation}
where $A_\mu$ is a 4-vector depending only on the light-cone
coordinate
\begin{equation}\label{eq:front}
    \xi = t -\frac{\boldsymbol{n}\cdot\boldsymbol{x}}{c}\,,
\end{equation}
with a unit constant 3-vector $\boldsymbol{n}$ pointing in the
direction of a wave propagation. Also we assume that the Coulomb
gauge is imposed which reduces to the condition
\begin{equation}\label{eq:gf}
    \boldsymbol{n}\cdot\boldsymbol{A}=0\,.
\end{equation}

Bellow it will be shown that the Lagrangian system (\ref{eq:nrL})
with the plane wave background of type (\ref{eq:plwave}) is
classically integrable and its solution can be represented in a
parametric integral form. Furthermore, after specialising to the
monochromatic plane wave background with an arbitrary elliptic
polarization, we derive the explicit form of the trajectory in
terms of the well-known Jacobian elliptic functions. To
demonstrate this we exploit the ideas from classical
Hamilton-Jacobi theory \cite{Whittaker,Arnold,AbrahamMarsden} as
well as the Dirac constraint formalism \cite{DiracLectures}.

%===========================================================================
\subsection{The Dirac parametrization trick and the Hamilton-Jacobi
equation}
%============================================================================

Due to the explicit time dependence of the electromagnetic wave
potential the Lagrangian (\ref{eq:nrL}) describes a  {\em
non-autonomous system}. It is convenient to enlarge the
corresponding  configuration space in such a way that the resulting
extended system becomes autonomous at the expense of being
invariant under an arbitrary time parametrization. This method is
often used in classical mechanics (see, for example, \cite{Arnold},
page 90 or \cite{AbrahamMarsden}, page 235) and also known to
particle physicists as the Dirac ``parametrization trick''
\cite{DiracLectures,HenneauxTeitelboim}.

The basic elements of this approach are the following. Starting
from an arbitrary Lagrangian system with Lagrangian
$\mathcal{L}\big( \boldsymbol{x}(t)\,,
\mathrm{d}\boldsymbol{x}/\mathrm{d} t\,,{t}\big)$ the
configuration space is extended  by considering time $t$ as a new
dynamical variable $t(s)$, which, together with the other
``spatial'' coordinates $\boldsymbol{x}(s)\,,$ depends upon the
auxiliary evolution parameter $s\,.$  The dynamics of the extended
system is determined from the degenerate, homogeneous (degree
one), time-reparametrization invariant Lagrangian
$\mathcal{L}^\ast$ constructed from the initial $\mathcal{L}$
according to the rule:
\begin{equation}\label{eq:repinvLagr}
\mathcal{L}^\ast\bigg(\boldsymbol{x}(s)\,,
 {t}(s)\,, \dot{\boldsymbol{x}}(s)\,,
 \dot{t}(s)\bigg):=\left(\frac{\mathrm{d}t}{\mathrm{d}s}\right)\mathcal{L}\bigg(
\boldsymbol{x}(s)\,, \frac{\mathrm{d}\boldsymbol{x}}{\mathrm{d}s}/
\frac{\mathrm{d}t}{\mathrm{d}s}\,,\, {t}(s)\bigg)\,.
\end{equation}
Hereafter a ``dot''  over a variable denotes a derivative with
respect to the evolution parameter $s\,$ and we require that
$t(s)$ is a monotonic, increasing function of the new evolution
parameter $s\,$
\begin{equation}\label{eq:mon}
    \frac{\mathrm{d}t}{\mathrm{d}s} >0\,.
\end{equation}
Numerically the new classical action based on the extended
Lagrangian (\ref{eq:repinvLagr}) is the same as the action of the
initial system, however,  it  turns out to be invariant with respect
to an arbitrary monotonic change of the evolution parameter
\begin{equation}\label{eq:repch}
    s \to s'=f(s)\,.
\end{equation}
 The extended phase of the system (\ref{eq:repinvLagr})
consists of $3+1$ canonical pairs
\begin{equation}\label{eq:cps}
 \boldsymbol{Z}(s):=\left[\begin{array}{cc}
  \boldsymbol{x}(s)\,, & \boldsymbol{p}(s) \\
     t(s)\,, & p_t(s)
  \end{array}
\right]\,, \qquad\ \boldsymbol{p}:=\frac{\partial\mathcal{L}^\ast
}{\partial\dot{\boldsymbol{x}}(s)}\,, \qquad\
p_t:=\frac{\partial\mathcal{L}^\ast }{\partial\dot{t}(s) }\,,
\end{equation}
but  due to the parametrization invariance  (\ref{eq:repch}) the
dynamics are constrained to develop on the surface of the phase
space defined by the equation
\begin{equation}\label{eq:constt}
  \mathcal{H}:= p_t+H_c = 0\,,
\end{equation}
where  $H_c$ is the canonical Hamiltonian corresponding to the initial
Lagrangian $\mathcal{L}\,.$

According to the Hamilton-Dirac description
\cite{DiracLectures,HenneauxTeitelboim} the constraint
(\ref{eq:constt}) plays a twofold role. First, it is a generator
of the local symmetry transformation of the phase space
coordinates (\ref{eq:cps}) induced by the time reparametrization
(\ref{eq:repch}). Second, inasmuch as the canonical Hamiltonian
derived from the Lagrangian $\mathcal{L}^\ast$ is identically zero
the dynamics are encoded in the constraint (\ref{eq:constt}) also.
This constraint generates  the evolution of the extended system
via the Hamilton-Dirac equations
\begin{equation}\label{eq:HDmotion}
    \dot{\boldsymbol{Z}}=\lambda(s)\{\boldsymbol{Z}\,,\mathcal{H}\}\,,
\end{equation}
with an arbitrary function $\lambda(s)\,$. The arbitrariness of
the Lagrange multiplier $\lambda(s)$ reflects the freedom to use an arbitrary
evolution parameter. Fixing it,  by imposing an
additional constraint
$$\chi(s, \boldsymbol{x}, t)=0\,,$$
allows one to show that any solution to (\ref{eq:HDmotion}) either
coincides with the classical trajectory of the initial Lagrangian
$\mathcal{L}\,,$ if gauge $\chi:=t-s$ is chosen,  or by using any
other admissible gauge, is canonically equivalent to it
\cite{HenneauxTeitelboim}.

Applying this general scheme to (\ref{eq:nrL}) the Lagrangian is
transformed to
\begin{equation} \mathcal{L}^{\ast}\left(\boldsymbol{x}\,,
    \dot{\boldsymbol{x}}\,,
    t\,, \dot{t}\right) =
    \frac{m}{2}\,\left( \frac{\dot{\boldsymbol{x}}}{\dot{t}}\right)^2\,
    \dot{t} \, +
    \frac{e}{c}\,\dot{\boldsymbol{x}}\cdot \boldsymbol{A}(\xi) -
    e \,\dot{t} \, \Phi(\xi) \,
    \end{equation}
and the time reparametrization invariant Hamiltonian dynamics of
the non-relativistic particle is  governed by the following
Hamiltonian constraint
\begin{equation}\label{eq:1pconstr}
   \mathcal{H}:= p_t+ e\Phi + \frac{1}{2m}
    \bigg(\boldsymbol{p}-\frac{e}{c}\,
    \boldsymbol{A}\bigg)^2
    \,=\,0\,.
\end{equation}

Our plan  to find the trajectory of a non-relativistic
particle in an electromagnetic plane background is the following. At
first, exploiting the Hamilton-Jacobi method, we perform a
canonical transformation to a free theory.  Having the explicit
form of this canonical transformation as well as the solution to
the free equations of motion we will be able to write down the
solution to the Hamilton-Dirac equation of motion
(\ref{eq:HDmotion}). Then, fixing with a suitable gauge, the Lagrange
multiplier function $\lambda(s)$ will be determined and finally
the trajectories of a charged particle will be given in a
parametric form analogous to the corresponding relativistic
problem.

%===================================================================
\subsection{Canonical transformation to a free system}
%===================================================================

 Following the basic idea of the Hamilton-Jacobi method consider a canonical transformation
\begin{equation}\label{eq:cantr}
\boldsymbol{Z}=\left[\begin{array}{cc}
  \boldsymbol{x}(s)\,, & \boldsymbol{p}(s) \\
     t(s)\,, & p_t(s)
  \end{array}
\right]\, \qquad\, \longleftrightarrow\,\qquad\,
\boldsymbol{Z}_0=\left[\begin{array}{cc}
  \boldsymbol{X}(s)\,, & \mathbf{\Pi}(s) \\
     T(s)\,, & \Pi_T(s)
  \end{array}
\right]\,,
\end{equation}
that  turns the constraint (\ref{eq:1pconstr}) into the constraint
of a free theory
\begin{equation}\label{eq:trenconstr}
   \mathcal{H}_0 = \Pi_T+ \frac{1}{2m}\,{\mathbf{\Pi}}^2=0\,.
\end{equation}
This  canonical transformation ``absorbs '' the electromagnetic
field and as a result the constraint $\mathcal{H}_0$ generates
according to (\ref{eq:HDmotion}) the simple free evolution:
\begin{eqnarray}\label{eq:sol1}
T(s)=T_{0}+\int_{0}^s\mathrm{d}u\, \lambda(u)\,, \qquad
\boldsymbol{X}(s)=
\boldsymbol{X}_{0}+\frac{\mathbf{\Pi}}{m}\,\int_{0}^s\mathrm{d}u\,
\lambda(u)\,.
\end{eqnarray}
In this solution  $\Pi_T\,$ and  $\mathbf{\Pi}$ denote constants
of motion which contain all the information about the initial
position  and velocity of the particle when $ s=0\,.$ The knowledge of
the explicit form of the ``absorbing transformation''
$\boldsymbol{Z} \to \boldsymbol{Z}_0$ is equivalent to solving
the initial interacting problem and can be found in exceptional
cases only. Fortuitously, it is the case for the problem we are
considering here.

The  ``absorbing'' transformation (\ref{eq:cantr}) can be
established  within the well-known method of generating function
\cite{Whittaker,Arnold} using  the  $S_2$\--function of old
coordinates $(t, \boldsymbol{x})$ and new momenta $(\Pi_T,
{\mathbf{\Pi}})$ of the form
\begin{equation}\label{eq:genf}
   {S}_2(t, \boldsymbol{x}, \Pi_T, {\mathbf{\Pi}})= t\,\Pi_T +
    \boldsymbol{x}\cdot\mathbf{\Pi}+ \mathcal{F}(\xi, \mathbf{\Pi})\,,
\end{equation}
with the unknown function $ {\mathcal{F}}(\xi, \mathbf{\Pi})\,$ to
be determined as follows.  Write the old momenta $p_t$ and
$\boldsymbol{p}$ as function of transformed coordinates
\begin{eqnarray}\label{eq:momA}
  p_t &=& \frac{\partial {S}_2}{\partial t}
  =\Pi_T + \frac{\mathrm{d} \mathcal{F}}{\mathrm{d}\xi}\,, \\
  &&\nonumber\\
  \boldsymbol{p}&=& \frac{\partial {S}_2}{\partial \boldsymbol{x}}=
  \mathbf{\Pi}\,-\frac{\boldsymbol{n}}{c}\,\frac{\mathrm{d}
  \mathcal{F}}{\mathrm{d}\xi}\,.\label{eq:momB}
\end{eqnarray}
Decompose all 3-vectors into their orthogonal and parallel
components with respect to the direction of wave propagation, e.g.,
$
    \mathbf{\Pi}= \mathbf{\Pi}_\bot +
    \Pi_\parallel\,\boldsymbol{n}\,,$
and by using the gauge condition (\ref{eq:gf}) we see that the
constraint (\ref{eq:1pconstr}) reduces to a free constraint
(\ref{eq:trenconstr}) if the function $\mathcal{F}$ is a solution
to the equation
\begin{equation}\label{eq:FnA}
   \biggl[ \frac{1}{c}\,\frac{\mathrm{d}\mathcal{F}}{\mathrm{d}\xi} +
   (mc-\Pi_\parallel)\biggl]^2=
   (mc-\Pi_\parallel)^2+ W(\xi, \mathbf{\Pi}_\bot)\,,
\end{equation}
where
\begin{equation}\label{eq:w}
    W(\xi, \mathbf{\Pi}_\bot):=-\frac{e^2}{c^2}\,\boldsymbol{A}^2_\bot
    +2\,\frac{e}{c}\,\boldsymbol{A}_\bot\cdot\mathbf{\Pi}_\bot+2me\,\Phi\,.
\end{equation}
The left hand side of (\ref{eq:FnA}) is positive definite,
therefore, a solution to (\ref{eq:FnA}) is real function  if
\begin{equation}\label{eq:noeqconA}
(mc-\Pi_\parallel)^2+ W(\xi, \mathbf{\Pi}_\bot)\geq 0\,.
\end{equation}
When the condition (\ref{eq:noeqconA}) is satisfied and imposing the
boundary condition $\mathcal{F}(0)=0\,,$ we have the solution:
\begin{equation}\label{eq:solF}
  \mathcal{F}(\xi\,, \mathbf{\Pi}) = -c(mc-\Pi_\parallel)\,\xi +
   c\int_0^\xi \mathrm{d}u\,
 \sqrt{(mc-\Pi_\parallel)^2+ W(u, \mathbf{\Pi}_\bot)}\,.
\end{equation}
 Here we insist that the inequality
(\ref{eq:noeqconA}) is satisfied  identically for all values of
integration variable $u$.  The importance of the restriction
(\ref{eq:noeqconA}) will be discussed later by analyzing the special
case of a monochromatic wave background.

Now substituting  the function $\mathcal{F}$ from (\ref{eq:solF})
into  (\ref{eq:momA}) and  (\ref{eq:momB}) one can easily
determine the expressions for a new momenta as function of the
initial ones. The momentum which is canonically conjugated  to the new time
coordinate $T $ is
\begin{eqnarray}\label{eq:ntmomentum}
 \Pi_T &=& p_t-c\biggl[(mc-p_\parallel)+
 \sqrt{(mc-p_\parallel)^2 - W(\xi\,, \boldsymbol{p}_\bot)}\, \bigg]\,,
\end{eqnarray}
and the new three dimensional momenta read
\begin{eqnarray}\label{eq:nxmomenta}
  \mathbf{\Pi}_\bot=\boldsymbol{p}_\bot\,,
  \quad
 \Pi_\parallel = mc+\sqrt{(mc-p_\parallel)^2- W(\xi\,,
 \boldsymbol{p}_\bot)} \,.
\end{eqnarray}

Using the generating equations
\begin{eqnarray}\label{eq:nec}
  T &=&
  \frac{\partial S_2}{\partial \Pi_T}=t\,,\\
  &&\nonumber\\
  \boldsymbol{X}&=& \frac{\partial S_2}{\partial \boldsymbol{\Pi}}=
  \boldsymbol{x}+ \frac{\mathrm{\partial} \mathcal{F}}{\partial
  \boldsymbol{\Pi}}\,
  \biggl|_{\,\mathbf{\Pi}=\mathbf{\Pi}(t, \boldsymbol{x},
  \boldsymbol{p})}\,,\label{eq:nec2}
\end{eqnarray}
one can find the new coordinates as a function of the old ones.
According to these results the time coordinate is unchanged
\begin{equation}\label{eq:newtime}
    T=t\, ,\\
\end{equation}
while the new three dimensional coordinates are
\begin{eqnarray}\label{eq:newcoordinates1}
    \boldsymbol{X}_{\bot}&=&\boldsymbol{x}_{\bot} +
   \frac{e}{|mc-p_\parallel|}
    \int_{0}^{\xi}
\, \mathrm{d}u\ \boldsymbol{A}_{\bot}(u)\,,\\
 {X}_{\|
 }&=& {x}_{\|
 }+ c\xi +
   \frac{c}{|mc-p_\parallel|}
 \int_{0}^{\xi}
    \, \mathrm{d}u\ \sqrt{\big(mc-p_\parallel\big)^2 -
    W(u, \boldsymbol{p}_\bot)}\,.\label{eq:newcoordinates2}
\end{eqnarray}

Equations (\ref{eq:nxmomenta}) express three independent constants
of motion, the two first constants coincide with the transversal
momenta $\boldsymbol{p}_\bot\,,$ while the third one
$\Pi_\parallel\,, $ can be interpreted as the longitudinal momenta
of particle only in the asymptotic region where the interaction
with the electromagnetic field is negligible\footnote{For further
interpretation see equations (\ref{eq:const}) and
(\ref{eq:constvz}) below. }. Note also, from (\ref{eq:ntmomentum})
and (\ref{eq:nxmomenta}), it follows that the so-called {\it
``light-cone energy''} represents a constant of motion
\begin{equation}\label{eq:lcenergy}
    \frac{p_t}{c} +p_\parallel=
    \frac{\Pi_T}{c} +\Pi_\parallel= \ constant\,.
\end{equation}

Suppose now that we are able to invert equations
(\ref{eq:newcoordinates1})-(\ref{eq:newcoordinates2}), i.e.
express the old canonical pairs $(t\,,p_t)$ and
$(\boldsymbol{x}\,,\boldsymbol{p})$ as functions of $(T\,,\Pi_t)$
and $(\boldsymbol{X}\,,\mathbf{\Pi})\,.$ Further, since
$(T\,,\Pi_t)$ and $(\boldsymbol{X}\,,\mathbf{\Pi})$ are known from
(\ref{eq:sol1}) (up to a gauge fixing),  these inverted
expressions give the solution to the Hamilton-Dirac equations
(\ref{eq:HDmotion}). To succeed in this inversion it is necessary to impose an
appropriate gauge condition on the
coordinates $t$ and $\boldsymbol{x}$.

%===================================================================
\subsection{Light-cone  gauge fixing and parametric solution}
%===================================================================

The observation that the light cone energy (\ref{eq:lcenergy}) is
a constant of motion  suggests a natural gauge fixing condition.
Namely, the evolution parameter $s$ can be identified with the
canonical variable conjugated to the light cone energy
\begin{equation}\label{eq:gfix}
  \chi:= t(s)-\frac{x_\parallel(s)}{c} - s=0\,.
\end{equation}

To find the Lagrange  multiplier function $\lambda\,,$ note that
according  to  the solution (\ref{eq:sol1})
\begin{equation}T - \frac{X_\parallel}{c} = (1-\frac{\Pi_\parallel}{mc})
\, \int_0^s \,du \, \lambda(u) \,,
\end{equation}
where the initial conditions $T_0$ and $\boldsymbol{X}_0$ have
been set equal to zero for simplicity. Furthermore, by using this
relation  and the equations
(\ref{eq:newtime})-(\ref{eq:newcoordinates2}) with  the gauge
fixing condition (\ref{eq:gfix}) implemented,  we find that the
Lagrange  multiplier obeys the following integral relation
\begin{equation}\label{eq;LMREl}
\int_0^s \,du \, \lambda(u) = \frac{mc}{\omega}\,\int_0^{\omega
s}\, du \, \frac{1}{\sqrt{(\Pi_\parallel- mc)^2 + W(u\,,
\boldsymbol{\Pi}_\bot)}}\,.
\end{equation}

With the aid of (\ref{eq;LMREl}) the equations
(\ref{eq:newtime})-(\ref{eq:newcoordinates2}) can be inverted with
respect to the initial coordinates
\begin{eqnarray}
\label{eq:1nparsolution}
  t(s) &=& mc\int_0^{s} \mathrm{d}u\,
 \frac{1}{\sqrt{(\Pi_\parallel-mc)^2+W(u,\mathbf{\Pi_\bot}
 )}}\,, \\ \label{eq:2nparsolution}
  x_\parallel(s) &=& -cs + mc^2\int_0^{s} \mathrm{d}u\,
 \frac{1}{\sqrt{(\Pi_\parallel-mc)^2+W(u,\mathbf{\Pi_\bot}
 )}}\,, \\
 &&\nonumber\\
  \boldsymbol{x}_\bot(s)&=& c\int_0^{s} \mathrm{d}u\,
 \frac{\mathbf{\Pi_\bot}-\displaystyle\frac{e}{c}\,
 \mathbf{A_\bot}(u)}{\sqrt{(\Pi_\parallel-mc)^2+W(u,\mathbf{\Pi_\bot}
 )}}\,.\label{eq:3nparsolution}
\end{eqnarray}
The formulae (\ref{eq:1nparsolution})-(\ref{eq:3nparsolution})
gives the parametric solution for a non-relativistic particle's
trajectory in an arbitrary plane wave background. They are in
a close analogy with the  parametric solution of  the corresponding
relativistic problem \cite{LandauLifshitz,ItzyksonZuber,Thirring}.

In order to write down  the three dimensional trajectory
$\boldsymbol{x}(t)\,,$ as a function of the physical time,  it is
necessary to invert (\ref{eq:1nparsolution}), i.e., to find $s$ as
function of $t\,$. It is  not possible to write down  an explicit
formula for an arbitrary plane wave background. However, in the
Section~\ref{sec3}, it will be shown how to solve this problem for
a monochromatic plane wave.

%%%%%%%%%%%%%%%%%%%%%%%%%%%%%%%%%%%%%%%%%%%%%%%%%%%%%%%%%%%%%%%%%%%%%%%%%%%%%%%%%
\subsection{A particle orbit in a weak plane wave background
}\label{sec:weakplane}
%%%%%%%%%%%%%%%%%%%%%%%%%%%%%%%%%%%%%%%%%%%%%%%%%%%%%%%%%%%%%%%%%%%%%%%%%%%%%%%%%%%

Before considering the inversion of the integral
(\ref{eq:1nparsolution}) for a monochromatic plane wave we will
sketch the possibility to solve the same problem for an arbitrarily
``weak'' plane wave in the form of and expansion in the
intensity parameter.

According to (\ref{eq:1nparsolution})-(\ref{eq:3nparsolution}), if
$s$ as a function of $t$ is known, $ s=f^{-1}(t)\,,$ the classical
trajectory can written in the form of the integral
\begin{equation}\label{eq:xex}
\boldsymbol{x_\bot}(t)=\displaystyle\frac{\boldsymbol{\Pi_\bot}}{m}\,t-
\displaystyle\frac{e}{mc}\,\int_{0}^{t}\mathrm{d}t'\,\boldsymbol{A_\bot}
(f^{-1}(t'))\,,
\end{equation}
and
\begin{equation}\label{eq:zex}
z(t)= c t- c f^{-1}(t)\,,
\end{equation}
These  formulae, with the function $f^{-1}$ determined from
(\ref{eq:1nparsolution}), gives a non-relativistic particle's
trajectory as function of LAB frame time in an arbitrary plane
wave background.

A ``naive'' non-relativistic limit of the solution for a particle's
trajectory follows from (\ref{eq:xex}) and (\ref{eq:zex})
by assuming the validity of the formal $1/c$ expansion of the
denominator of the integrand in the expression
(\ref{eq:1nparsolution}). In this case, for small enough laser
intensities,  a closed form of the charged particle's classical
trajectory, $\boldsymbol{x}(t),$ as function of the LAB frame time
$t$ can be written straightforwardly.

Indeed, in the approximation $mc-\Pi_\parallel \approx mc\, $ and $
\boldsymbol{\Pi_\bot}/mc \approx 0\,, $ keeping only the leading
term of the $1/mc$-expansion of the denominator in
(\ref{eq:1nparsolution}) we have
\begin{eqnarray}\label{eq:wexp}
  t(s) = s+\frac{1}{2}\,\eta^2\int_0^{s } \mathrm{d}u\,
\boldsymbol{a}^2_\bot(u)\,.
\end{eqnarray}
In (\ref{eq:wexp}) the normalized potential  \(
\boldsymbol{a}_\bot:= \boldsymbol{A}_\bot/\sqrt{\langle
\boldsymbol{A}^2_\bot \rangle} \, \) with  $\langle \dots \rangle$
denoting the time average  and the dimensionless intensity
parameter $\eta $,
\begin{equation}\label{eq:eta}
\eta^2=-2\,\frac{e^2}{m^2c^4}\,\langle A_\mu\,A^\mu\rangle\,,
\end{equation}
have been introduced.

Therefore, for small intensities,  the auxiliary time  $s$  in the
leading $\eta$ order is
\begin{equation}\label{eq:st}
    s=t-
\frac{1}{2}\,\eta^2\int_0^{t} \mathrm{d}u\,
\boldsymbol{a}^2_\bot(u)+\dots\,,
\end{equation}
and the approximate form of a  charged particle's trajectory reads
\begin{equation}\label{eq:xexexpans}
\boldsymbol{x_\bot}(t)=
\displaystyle\frac{\boldsymbol{\Pi_\bot}}{m}\,t-
c\eta\,\int_{0}^{t}\mathrm{d}u\,\boldsymbol{a_\bot} (u) +
\displaystyle\frac{\boldsymbol{\Pi_\bot}}{2m}\,\eta^2\int_0^{t}
\mathrm{d}u\, \boldsymbol{a}^2_\bot(u) +\dots \,,
\end{equation}
\begin{equation}\label{eq:zexapr}
z(t)= \frac{1}{2}\,c\eta^2\int_0^{t} \mathrm{d}u\,
\boldsymbol{a}^2_\bot(u)+\dots\,.
\end{equation}

The higher order corrections can be obtained in a similar way
using, for example, the well-known Lagrange expansion method
 over the small parameter \cite{WhittakerWatson}.

%%%%%%%%%%%%%%%%%%%%%%%%%%%%%%%%%%%%%%%%%%%%%%%%%%%%%%%%%%%%%%%%% III
\section{Particle's orbit in a monochromatic plane wave}\label{sec3}
%%%%%%%%%%%%%%%%%%%%%%%%%%%%%%%%%%%%%%%%%%%%%%%%%%%%%%%%%%%%%%%%%%%%%%

In this section we exploit the generic parametric representation
for a particle's trajectory found above for the practically
important case of a charge's propagation in the background of a
monochromatic plane wave with an arbitrary polarization.

We specify the gauge potential in equations
(\ref{eq:1nparsolution})\,-\,(\ref{eq:3nparsolution}) as
\begin{equation}\label{eq:gpot}
    A^\mu : = a(u)\,\biggl(0\,, \ \varepsilon\cos(u)\,,\
    \sqrt{1-\varepsilon^2}\sin(u)\,,\ 0
    \biggl)\,, \qquad u =\omega_L\left(t-\frac{z}{c}\right)\,.
\end{equation}
Here, the four vector in brackets describes a monochromatic
elliptically polarized electromagnetic plane wave with frequency
$\omega_L$ travelling in the $z$-direction.  The parameter
$0\leq\varepsilon\leq 1$ measures the polarization in a way such
that  the boundary values $\varepsilon=0$ and $\varepsilon=1$
correspond to linear polarization, while
$\varepsilon=1/\sqrt{2}\,$ to circular polarization. In
application to laser beams the profile function $a(u)$ is usually
assumed to be smooth and slowly varying (on the scale of
oscillations) and vanishing at $u \to \pm \infty$. In this article
we are not intending to discuss a realistic laser and therefore, for the
remainder of calculation, the pulse function is
chosen to be constant $a(u):= a $.  Formally this corresponds to a
laser with an infinite length pulse.

Any solution to the classical equation of motion for a particle
depends on the laser field's characteristics as well as on the
initial/boundary conditions on a particle's position and velocity.
The laser field's characteristics used below are, the frequency
$\omega_L\,, $ the polarization $\varepsilon \,$  and the gauge
invariant dimensionless intensity parameter (\ref{eq:intensity})
which for our choice of monochromatic wave potential is
\begin{equation}\label{eq:etamon}
\eta^2=-2\,\frac{e^2}{m^2c^4}\,\langle A_\mu\,A^\mu\rangle=\,
\left(\frac{ea}{mc^2}\right)^2\,,
\end{equation}
where now $\langle\, \dots\, \rangle$ denotes the time averaging
over the period  $2\pi/\omega_L$.

The dependence of a particle's orbits on the initial/boundary
conditions is encoded via the first integrals
$\boldsymbol{\Pi}_\bot$ and $\Pi_\parallel\,.$  The analysis of a
generic boundary conditions is given in the
Appendix~\ref{ap:genericbound}. It is shown there how one can
invert equation (\ref{eq:1nparsolution}) expressing the auxiliary
evolution parameter $s$ in terms of the physical time $t$ via the
Weierstrass elliptic function $\wp(\omega_L t)$. Having this
representation the remaining integrals (\ref{eq:2nparsolution})
and (\ref{eq:3nparsolution}) with the arbitrary constants
$\boldsymbol{\Pi}_\bot$ and $\Pi_\parallel\,$ determine the
explicit form of a particle's trajectory as function of the physical
time $t\,.$ However, to make the solution more transparent, we omit this rather
technical work here and prefer to describe the orbits
for a restricted  but nevertheless informative initial conditions.

Recall that the solution
(\ref{eq:1nparsolution})-(\ref{eq:3nparsolution}) is written when
the initial conditions on the coordinates (in physical time $t$)
read
\begin{equation}\label{eq:initialcond}
\boldsymbol{x}(t=0) = 0\,,
\end{equation}
and the initial velocity $\boldsymbol{\upsilon}(0):=
\mathrm{d}\boldsymbol{x}/\mathrm{d}t(t=0)\,$ is expressible via
the dimensionless constants of motion
\begin{equation}\label{eq:dimpara}
   \beta_+ :=1-\frac{\Pi_\parallel}{mc}\,,\qquad
\boldsymbol{\beta}_{\bot}=(\beta_1,
\beta_2):=\frac{\boldsymbol{\Pi_{\bot}}}{mc}\,,
\end{equation}
with the aid of  relations
\begin{eqnarray}\label{eq:const}
\boldsymbol{\upsilon}_\bot(0)&=&c \boldsymbol{\beta}_\bot -c\eta\,
\boldsymbol{\epsilon}_\bot\,,\\
\upsilon_z(0)&=&c-
c\,\sqrt{\beta_+^2-\eta^2\varepsilon^2+2\,\eta\,
\boldsymbol{\epsilon}_\bot\cdot\boldsymbol{\beta}_\bot\,}\,,\label{eq:constvz}
\end{eqnarray}
where for  the choice (\ref{eq:gpot}) we have
$\boldsymbol{\epsilon}_\bot=(\varepsilon, 0)\,.$ Since for small
velocities the system possesses a Galilean symmetry, (see Appendix
\ref{sec:appendix1})  we can pass to a certain reference frame by
specifying the constants $\boldsymbol{\upsilon}(0)\,$. Below we
restrict ourselves by insisting the vanishing of the
transverse velocity\footnote{As it will be shown below the
fixation $\boldsymbol{\beta}_\bot=0\,,$ corresponds to zero
average transverse velocity,
$\langle\boldsymbol{\upsilon}_\bot\rangle =0\,.$ }
\begin{eqnarray}\label{eq:zerotransv}
\boldsymbol{\beta}_\bot = 0\,.
\end{eqnarray}
Therefore, the orbits are specified by the constant,
\begin{equation}\label{eq:longvel}
    \beta_z=\frac{\upsilon_z(0)}{c}=
1-\sqrt{\beta_+^2-\eta^2\varepsilon^2\,}\,,
\end{equation}
characterizing the particle's longitudinal velocity at $t=0\,.$ In
order to have a  real velocity we require $\beta_+^2
> \eta^2\varepsilon^2\,.$

Now with this specific choice of constants,  we shall find $x(t),\,
y(t)$ and $z(t)$ as functions of the physical time $t\,.$ Here we also note that
due to the $2\pi$- periodicity of the monochromatic
gauge potential  (\ref{eq:gpot}) all canonical coordinates will be
treated as functions of a point on a circle with the standard
trigonometric parametrization restricted to the principle domain $
[-\pi/2 \,, \pi/2 ] \,.$

%===================================================================
\subsection{Fundamental domain and fundamental solution}
%===================================================================
\label{subsec3.1}

 Now we want to describe a particle's trajectory as a function
of the laboratory frame's time $t$.  First of all let us preserve the
condition (\ref{eq:noeqconA}) which guaranties the monotonic
character of the function $t(s)$  (see equation (\ref{eq:mon})), as well as
reality of particle's trajectories is satisfied. For the monochromatic plane wave
(\ref{eq:gpot}) the inequality (\ref{eq:noeqconA}) with the
vanishing transverse momentum, $\boldsymbol{\Pi}_\bot = 0\,,$ can
be rewritten as
\begin{equation}\label{eq:noeqconmon}
1-\mu^2\sin^2u  > 0\,,
\end{equation}
where
\begin{equation}\label{parameter}
\mu^2: = (1-2\,\varepsilon^2)\,\frac{\eta^2}{(1-\beta_z)^2}\,.
\end{equation}
We define three  allowed domains as
\begin{equation*}
\mathrm{(I)}\quad   0 <\mu^2 < 1\,,\quad \mathrm{(II)}\qquad
\mu^2
> 1\,,\qquad \mathrm{(III)}\quad \mu^2 < 0\,.
\end{equation*}
Below we show that having the knowledge of the solution for region (I),
which we call the {\em ``fundamental domain''}, determines the
solutions in all other regions. Namely, it will be demonstrated
that all possible particle trajectories can be obtained from the
{\em ``fundamental solutions''} (solution depending on parameters
from the ``fundamental domain'') by combination of inversion $\mu
\to 1/\mu$, and rotation to the imaginary axis $\mu \to \imath
\mu\,.$

Besides this, we also present solutions for two special cases
$$\mu^2=0\, \qquad \mathrm{and}\qquad \mu^2=1\,.$$
We emphasize  here that these solutions can also be described by
starting from the orbits from the fundamental domain and taking a
corresponding limit.

To prove the above statements we  start with the analysis  of the
``fundamental solutions'' and derive  a particle's trajectory in
terms of the physical time $t\,.$

%^^^^^^^^^^^^^^^^^^^^^^^^^^^^^^^^^^^^^^^^^^^^^^^^^^^^^^^^^^^^^^^^^^^^^^^^
\subsubsection{Orbits for the fundamental domain,
$(\mathrm{I}):\,$ $ 0<\mu^2<1\,.$}
%^^^^^^^^^^^^^^^^^^^^^^^^^^^^^^^^^^^^^^^^^^^^^^^^^^^^^^^^^^^^^^^^^^^^^^^^^^^^^^^^^^^

For the parameters from the ``fundamental domain'' (I) the
inequality (\ref{eq:noeqconmon}) is true for all values of $s$
from the interval $ -\pi/2 \leq u \leq \pi/2\,.$ Equations
(\ref{eq:1nparsolution}), (\ref{eq:3nparsolution}) can be
rewritten as
\begin{eqnarray}
\label{eq:1monsolution}
  t(s) &=& \frac{1}{\omega_L (1-\beta_z)}\,\int_0^{\omega_L s } \mathrm{d}u\,
 \frac{1}{\sqrt{1-\mu^2\,\sin^2u }}\,, \\
  x(s) &=&
  -\frac{c}{\omega_L}\,\sqrt{\frac{\varepsilon^2}{1-2\varepsilon^2}}
  \,\arcsin \left[ \sqrt{\mu^2}\,\sin(\omega_L s)\right]\,,
  \label{eq:1monsolutionX}\\
  y(s)&=& \frac{c}{\omega_L }\,\sqrt{\frac{1-\varepsilon^2}{1-2\,\varepsilon^2}}
  \,\ln
  \left[\frac{\sqrt{\mu^2}\,\cos(\omega_L s)+\sqrt{1-\mu^2\sin^2(\omega_L s)}}{1+\sqrt{\mu^2}}
  \right]
 \,,\label{eq:1monsolutiony}
\end{eqnarray}
and the component in the wave propagation direction is
\begin{equation}\label{eq:zcom}
z(s) =  c t(s) - c s \,.
\end{equation}
This is the parametric solution to the equation of motion with
parameter $s$ from the principal interval
\begin{equation}\label{eq:parrang}
 - \frac{\pi}{2} \leq {\omega_L}s \leq \frac{\pi}{2}\,.
\end{equation}
 Now, thanks to L.~Euler, A.M.~Legendre, N.H.~Abel and
C.G.J.~Jacobi, we know how to invert (\ref{eq:1monsolution}), i.e.
find the evolution parameter $s$  as function of the physical time
$t\,.$ This can be done in terms of the well-known Jacobian {\em
amplitude} function (\cite{WhittakerWatson,Bateman} and Appendix
\ref{sec:appendix2})
\begin{equation}\label{eq:s-t}
    \omega_L s=\mathrm{am}\big(\omega_L^\prime t,\,
    \mu\big)\,,
\end{equation}
with {\em modulus} $\mu:=\sqrt{\mu^2}\,$ and the non-relativistically
Doppler  shifted frequency
\begin{equation}\label{eq:Doppl}
    \omega_L^\prime:=\omega_L\,(1-\beta_z)\,.
\end{equation}

For the values of $s$ from the interval (\ref{eq:parrang}) the
amplitude function is a well defined increasing function defined on
the interval
\begin{equation}\label{eq:interint }
-\,\mathrm{K}(\mu)\,\leq \omega_L^\prime t \,\leq
\mathrm{K}(\mu)\,,
\end{equation}
where $\mathrm{K}$ is the ``real'' quarter period of the Jacobian
elliptic functions, (\ref{eq:KkK'}). Therefore, one can consider the
transformation from the evolution parameter $s$ to time $t$ as
well-defined change of coordinates on a circle.

Substituting  the expression  for the evolution parameter in terms
of the physical time (\ref{eq:s-t}) into  (\ref{eq:1monsolutionX}),
(\ref{eq:1monsolutiony}) and (\ref{eq:zcom}) and  using the
properties of the  Jacobian functions we arrive at the
representation of the classical trajectory
\begin{eqnarray}\label{trajectoryLAB}
x_{\mathrm{F}}(t) &=&
  -\frac{c}{\omega_L}\,\sqrt{\frac{\varepsilon^2}{1-2\,\varepsilon^2}}
  \,\arcsin \left[\mu\,
  \mathrm{sn}\big(\omega_L^\prime t,\, \mu\big)\right]\,, \\
  y_{\mathrm{F}}(t)&=& \frac{c}{\omega_L }\,\sqrt{\frac{1-\varepsilon^2}{1-2\,\varepsilon^2}}
  \,\ln \left[
\frac{\mu\,\mathrm{cn}\big(\omega_L^\prime t,\, \mu\big)
+\mathrm{dn}\big(\omega_L^\prime t,\, \mu\big)}{1+\mu}
  \right]\,,
\end{eqnarray}
and
\begin{equation}\label{eq:zA}
z_{\mathrm{F}}(t) =  c t -
\frac{c}{\omega_L}\,\mathrm{am}\big(\omega_L^\prime t,\,
\mu\big)\,.
\end{equation}
Here the subscript ``F'' is written to emphasize that
(\ref{trajectoryLAB})-(\ref{eq:zA}) corresponds to the
trajectories with  the modulus from the fundamental interval
$0<\mu^2<1\,.$

Now we shall consider all the other domains of parameters and, using
the properties of the Jacobian functions, will give explicit
expressions for their corresponding trajectories.

%^^^^^^^^^^^^^^^^^^^^^^^^^^^^^^^^^^^^^^^^^^^^^^^^^^^^^^^^^^^^^^^^^^^^^^^^^^^^^^^^^^
\subsubsection{Orbits for the second domain,
$(\mathrm{II}):$ $\,\mu^2 > 1\,.$}
%^^^^^^^^^^^^^^^^^^^^^^^^^^^^^^^^^^^^^^^^^^^^^^^^^^^^^^^^^^^^^^^^^^^^^^^^^^^^^^^^^^

When analyzing this domain two peculiarities should be taken into
account. First of all, when  \( \, \mu^2
> 1\,\) the inequality (\ref{eq:noeqconmon})
is true only if
\begin{equation}
  \sin^2u <  \underline{\mu}^2\,, \qquad \underline{\mu}:=\frac{1}{{\mu}}\,.
\end{equation}
This means that  the upper limit to the integral in
(\ref{eq:1monsolution})  lies in the interval
\begin{eqnarray}\label{eq:sforg1}
-\frac{\pi}{2}<-\arcsin\left(\underline{\mu}\right)\leq &\omega_L
s&\leq \arcsin\left(\underline{\mu}\right)<\frac{\pi}{2}\,.
\end{eqnarray}

Second, since the standard integral representation of the
amplitude function (\ref{eq:amplitude}) is defined for a
modulus from the fundamental interval, $0 <\mu^2 < 1\,,$  some
simple mathematical manipulations are required to rewrite the integral
(\ref{eq:1monsolution}) in such a form that its integrand depends
on the inverse modulus,
 $\underline{\mu}\,,$ instead of modulus $\mu\,.$ Namely one can
 easily verify that (\ref{eq:1nparsolution}) can be written as
\begin{eqnarray}
\label{eq:1monsolutiong1}
  t(s) &=& \frac{1}{\omega_L^\prime{\mu}}\,
  \int_0^{ \arcsin(\mu\sin(\omega_L s) )} \mathrm{d}u\,
 \frac{1}{\sqrt{1-\underline{\mu}^2\,\sin^2u }}\,.
\end{eqnarray}
Therefore, the relationship between the evolution parameter $s$
and time $t$ is
\begin{equation}\label{eq:s-tII}
   \omega_L s
   =\arcsin \left(\underline{\mu}\,
   \mathrm{sn}\left(\omega_L^\prime{\mu}\,t,\,
   \underline{\mu}\right)\right)\,.
\end{equation}
When the parameter $s$ is contained in the interval (\ref{eq:sforg1})
relation (\ref{eq:s-tII}) defines an increasing function on the
interval
\begin{equation}\label{eq:interint2 }
-\mathrm{K}(\underline{\mu})\leq\omega_L^\prime{\mu}\,t
\leq\mathrm{K}(\underline{\mu})\,.
\end{equation}

Finally using (\ref{eq:s-tII})  and relations (\ref{jacobi:def})
we have
\begin{eqnarray}\label{trajectoryLABII}
x(t) &=&
  -\frac{c}{\omega}\,\sqrt{\frac{\varepsilon^2}{1-2\,\varepsilon^2}}
  \,\mathrm{am} \big(\omega_L^\prime{\mu}\,t,\,\underline{\mu}\big)\,, \\
 y(t)&=& \frac{c}{\omega }\,\sqrt{\frac{1-\varepsilon^2}{1-2\,\varepsilon^2}}
  \,\ln \left[
\frac{\,\mathrm{cn}\big(\omega_L^\prime{\mu}\,t,\,
\underline{\mu}\big)\, +\mu
\,\mathrm{dn}\big(\omega_L^\prime{\mu}\,t,\, \underline{\mu}
\big)}{\mu+1}
  \right]
 \,,
\end{eqnarray}
and
\begin{equation}\label{eq:zB}
z(t) =  c t - \frac{c}{\omega_L}\,\arcsin\left(\underline{\mu}\,
\mathrm{sn}\big(\omega_L^\prime{\mu}\,t,\,
\underline{\mu}\big)\right)\,.
\end{equation}
These formulae describe a particle's trajectory when the parameter
$\mu^2$  takes its value in the second domain, $\mu^2 > 1\,.$

%^^^^^^^^^^^^^^^^^^^^^^^^^^^^^^^^^^^^^^^^^^^^^^^^^^^^^^^^^^^^^^^^^^^^^^^^^^^^^^^^^^
\subsubsection{Orbits for the third domain, $(\mathrm{III}):$
 \,$\mu^2 < 0\,.$ }
%^^^^^^^^^^^^^^^^^^^^^^^^^^^^^^^^^^^^^^^^^^^^^^^^^^^^^^^^^^^^^^^^^^^^^^^^^^^^^^^^^^

If $\mu^2 < 0\,$ the inequality (\ref{eq:noeqconmon}) is true for
the whole  principal interval $ [-\pi/2\,, \pi/2]\,.$ Now
introducing the positive definite parameter $ \kappa^2>0\,,
{\mu^2}: =- \kappa^2\,,$ the relation (\ref{eq:1monsolution})
reads
\begin{eqnarray}\label{eq:imagrel}
  t(s) &=& \frac{1}{\omega_L^\prime}\,
  \int_0^{ws} \mathrm{d}u\,
 \frac{1}{\sqrt{1+\kappa^2\,\sin^2u }}\,,
\end{eqnarray}
 while the solution for the spatial coordinates is
\begin{eqnarray}
  x(s) &=&
  -\frac{c}{\omega'_L}\,\frac{\eta \varepsilon}{\kappa}
  \,\mathrm{arsinh} \left[ \kappa\,\sin(\omega_L s)\right]\,,
  \label{eq:1immonsolutionX}\\
  y(s)&=& \frac{c}{\omega'_L }\,
 \frac{ \sqrt{1-\varepsilon^2\,}}{i\kappa}\,\ln\left[
  \frac{\kappa\cos(\omega_Ls)+i\sqrt{1+\kappa^2\sin^2(\omega_Ls)}}{i+\kappa}
  \right]
 \,.\label{eq:1immonsolutiony}
\end{eqnarray}

Now again a change of integration variable must be done in
(\ref{eq:imagrel}) in order to get  the integral expressible in
terms of the Jacobian amplitude with  modulus from the open
interval \( (0,1)\), thereby guaranteing that all functions are
single-valued and continuous. It is straightforward to check that
(\ref{eq:imagrel}) is equivalent to
\begin{eqnarray} \label{eq:imagrel2}
 && t(s) = \frac{1}{\omega'_L\kappa'}\,
  \int_0^{\phi(s)}\mathrm{d}u\,
 \frac{1}{\sqrt{1-\displaystyle\frac{\kappa^2}{{\kappa'}^2}\,\sin^2u }}\,,
\end{eqnarray}
where the upper limit of the integral is
\begin{equation}\label{eq:uplimit}
    \phi(s): =\arcsin\left(
 \frac{\kappa^{\prime}\sin(\omega s)}{\sqrt{1+\kappa^2\sin^2(\omega s)}
 }\,
 \right)\,, \qquad {\kappa'}^{2} = 1+\kappa^2\,.
\end{equation}
We have now achieved the goal that the modulus of $\kappa/\kappa' \in
(0,1)\,, $ for all values $\mu^2 < 0\,.$ Therefore, the inverse to
(\ref{eq:imagrel2}) reads
\begin{equation}\label{eq:invri}
    \phi(s)= \mathrm{am}\left(\omega'_L \kappa't,\,
    \kappa/\kappa' \right)\,,
\end{equation}
or directly for the evolution parameter we get
\begin{equation}\label{eq:imgt-s}
  \kappa'\,\sin(\omega_L s) = \frac{ \mathrm{sn}\left(\omega'_L
    \kappa't,\,
    \kappa/\kappa' \right)}{ \mathrm{dn}\left(\omega'_L
    \kappa't,\,
     \kappa/\kappa' \right)}\,.
\end{equation}
With the aid of  (\ref{eq:imgt-s}) one can finally rewrite the
particle's trajectory in terms of the Jacobian elliptic function
with modulus ${\kappa}/{\kappa^\prime}$
\begin{eqnarray}
x(t) &=& -\frac{c}{\omega'_L}\,\frac{\eta\varepsilon}{\kappa}
\mathrm{arsinh}{\left[\displaystyle{\frac{\kappa}{\kappa^\prime}}\,\frac{
\mathrm{sn}\left(\omega'_L
    \kappa't,\,
    \kappa/\kappa' \right)}{ \mathrm{dn}\left(\omega'_L
    \kappa't,\,
     \kappa/\kappa' \right)}\right]}\,,\label{eq:imagtraj1}
\\
y(t)&=&
\frac{c}{\omega_L}\,\frac{\eta\sqrt{1-\varepsilon^2}}{i\kappa}\,
\ln{\left[ \frac{1-i\kappa\,\mathrm{cn}\left(\omega_L^\prime
\kappa^\prime\,t,\, \kappa/\kappa^\prime\right)}{(1-i\kappa)
\mathrm{dn}\left(\omega'_L \kappa't,\,\kappa/\kappa' \right)}
\right]}\,,\label{eq:imagtraj2}
\\
 z(t) &=&
 ct -\frac{c}{\omega_L}
\arcsin{\left[\frac{1}{\kappa'}\, \frac{
\mathrm{sn}\left(\omega'_L\kappa't,\,
    \kappa/\kappa' \right)}{ \mathrm{dn}\left(\omega'_L
    \kappa't,\,
     \kappa/\kappa' \right)}
\right]}\,.\label{eq:imagtraj3}
\end{eqnarray}

Equations (\ref{eq:imagtraj1})-(\ref{eq:imagtraj3}) describe a
particle's trajectory in the background characterized by $\mu^2< 0\,,$
and are  well defined on the interval
\begin{equation}
-\mathrm{K}(\kappa/\kappa') < \omega_L'\kappa' t <
\mathrm{K}(\kappa/\kappa') \,.
\end{equation}

%^^^^^^^^^^^^^^^^^^^^^^^^^^^^^^^^^^^^^^^^^^^^^^^^^^^^^^^^^^^^^^^^^^^^^^^^^^^^^^^^^^
\subsubsection{Degenerate orbits,  $\mu^2=0\,\  \&\, \ \mu^2=1\,.$}
\label{subsubsec:degen}
%^^^^^^^^^^^^^^^^^^^^^^^^^^^^^^^^^^^^^^^^^^^^^^^^^^^^^^^^^^^^^^^^^^^^^^^^^^^^^^^^^^

Here we consider the two special cases remaining from the previous
considerations. Namely, the special orbits with the parameter
$\mu^2=0$ and $\mu^2=1\, $ will be presented. Note that the
vanishing modulus corresponds to a special circularly polarized monochromatic plane wave,
$\varepsilon^2=1/2\,$ or the trivial zero background case.

If $\mu^2=0$, then the parametric solution
(\ref{eq:1nparsolution}) determining the time $t$ as  a function of an
auxiliary parameter $s$ takes the simple form
\begin{equation} \label{eq:timemu0}
  t(s) = \frac{1}{1-\beta_z}\,s\,,
\end{equation}
while the equations (\ref{eq:3nparsolution}) for the spatial
components  orthogonal to the direction of the wave propagation
reduce to
\begin{eqnarray}
  x(s) =- \frac{1}{\sqrt{2}}\,\frac{c}{\omega'_L}\,\eta\sin\omega_L s\,, \qquad
  y(s)= -\sqrt{2}\,\frac{c}{\omega'_L }\,\eta\,\sin^2\left(\frac{\omega_L s}{2}\right)\,.
\end{eqnarray}
Therefore, it follows from (\ref{eq:timemu0}), that for the
degenerate case of zero modulus $\mu^2=0$ the trajectory is
\begin{eqnarray*}\label{eq:orthmu0}
x(t) =
  -\frac{1}{\sqrt{2}}\,\eta\,\frac{c}{\omega'_L}\,\sin\omega_L^\prime t\,,
  \quad
  y(t)=-\frac{2}{\sqrt{2}}\,\eta\frac{c}{\omega'_L }\sin^2\left(\frac{\omega'_L t}{2}\right),
  \quad z(t) = \beta_z\,c\,t.\quad
\end{eqnarray*}
From these expressions we see that for a circular polarized
monochromatic wave all nonlinear effects disappear and one can
choose such a reference frame where the particle's motion appears as a
pure harmonic with the non-relativistically Doppler shifted laser
frequency $\omega_L^\prime\,.$

For the orbit  characterized by $\mu^2=1$, the equation
(\ref{eq:1nparsolution}) gives
\begin{equation}\label{eq:timemu1}
  t(s) = \frac{1}{\omega_L^\prime}\,\mathrm{artanh}\left(\sin\omega_L
s\right)\,,
\end{equation}
and from (\ref{eq:3nparsolution}) it follows that
\begin{eqnarray}\label{eq:xymu1}
  x(s) =
  - \sqrt{\frac{\varepsilon^2}{1-2\,\varepsilon^2}}\, cs\,,
\qquad
 y(s)= \sqrt{\frac{1-\varepsilon^2}{1-2\,\varepsilon^2}}\,\frac{c}{\omega_L} \ln\left(\cos\omega_L s\right)\,
 \,.
\end{eqnarray}

This finalizes our consideration of all possible particle's
trajectories in a generic monochromatic  plane wave.

In the next subsection we will briefly outline how all possible
particle's trajectories  can be categorized into the equivalent
classes with respect to the action of the
$SL(2,\mathbf{Z})/(1,-1)$ group.

%-----------------------------------------
\subsection{Modular properties of orbits}
%----------------------------------------------

As shown above, the nonlinear dependence on the polarization
and the intensity of the radiation background of a particle's
trajectory is encoded in the modulus of the elliptic Jacobian
functions. The doubly periodic elliptic functions have a
remarkable property related to a certain symmetry of modulus
transformations. Namely, the elliptic function with periods $w_1$
and $ w_2$ can be algebraically expressed through another elliptic
functions with periods $w'_1$ and $ w'_2$ if periods are related
by the so-called unimodular transformations
\begin{equation}\label{eq:modtr}
    \left(
      \begin{array}{c}
        w'_1 \\
        w'_2 \\
      \end{array}
    \right)
=\left(
   \begin{array}{cc}
     a & b \\
     c & d \\
   \end{array}
 \right)
\left(
  \begin{array}{c}
    w_1 \\
    w_2 \\
  \end{array}
\right)\,,
\end{equation}
where the entries of the $2\times2$  matrix are integers $a,b,c,d \in
    \mathbb{Z}\,$ and $ ad-bc=1\,. $
The transformations (\ref{eq:modtr}) may also be treated also as a
subgroup of the M\"{o}bius transformations in the upper half of the
complex $\tau $ plane into itself
\begin{equation}\label{eq:Moebius}
    \tau'=\frac{c+d\tau}{a+b\tau}\,,   \qquad \tau:=
    \frac{w_2}{w_1}\,.
\end{equation}
For the problem we are dealing with here this modular equivalence
exposes in the \textit{intensity duality}, i.e. there exist a
specific correspondence between the motion in backgrounds with
various intensity regimes. Any trajectory with an arbitrarily
prescribed intensity can be connected to  the solution from the
fundamental domain with the aid of  a certain modular
transformation. Particularly, if we assign to a particle's
trajectory with low intensity a certain modular parameter $\tau$,
then the trajectory for the high intensity conditions are related
by $\tau \to \tau/(1+\tau)\,. $ Indeed, using the relations
between the Jacobian functions whose moduli are inverse to each
other, (cf. formulae (\ref{eq:JacobitranC})), one can verify that
the expressions (\ref{trajectoryLABII})-(\ref{eq:zB}) with
$\mu^2>1$ follow from the fundamental solution
(\ref{trajectoryLAB})-(\ref{eq:zA}) by the substitution $\mu \to
1/\mu: $
\begin{equation}\label{eq:lhrel}
    \boldsymbol{x}(t\,|\,\mu)
    =\boldsymbol{x}_{\mathrm{F}}(t\,|\,\frac{1}{\mu}\, )\,,
\end{equation}
or in terms of the modular parameter $\tau $
\begin{equation}\label{eq:shifmod}
    \boldsymbol{x}(t\,|\,\tau)
    =\boldsymbol{x}_{\mathrm{F}}(t\,|\,\frac{\tau}{1+\tau}\,)\,.
\end{equation}
Analogously, if $\mu^2 < 0$, the trajectories
(\ref{eq:imagtraj1})-(\ref{eq:imagtraj3}) are connected to the
fundamental solution by the shift transformation
(\ref{eq:shifttran21}),
\begin{equation}\label{eq:shifmod2}
    \boldsymbol{x}(t\,|\,\tau)
    =\boldsymbol{x}_{\mathrm{F}}(t\,|{1+ \tau}\,)\,.
\end{equation}

Note also that the  special trajectories with $\mu^2=0$ and $\mu^2=1\,,$
considered above,  coincides with the corresponding limits of the
fundamental solution taking into account that the Jacobian
functions are degenerate to the trigonometric (\ref{eq:sndegmu2}) and
hyperbolic functions (\ref{eq:sndegmu1}) for moduli $\mu=0 \, $
and $\mu=1\,, $ respectively.

%%%%%%%%%%%%%%%%%%%%%%%%%%%%%%%%%%%%%%%%%%%%%%%%%% IV
\section{Analysis of the particle's trajectory }\label{sec4}
%%%%%%%%%%%%%%%%%%%%%%%%%%%%%%%%%%%%%%%%%%%%%%%%%%%%%%%%

Now we analyze in greater detail the trajectory with parameters
from the fundamental domain and clarify some of the physical features of
this solution.

From the solution (\ref{trajectoryLAB})-(\ref{eq:zA}) and (\ref{jacobi:derivatives})-(\ref{jacobi:derivatives2}) one can
derive an expression for the particle's velocity
\begin{eqnarray}\label{eq:xvelocity}
\upsilon_x(t)&=&-c\eta\epsilon\,\mathrm{cn}\left(\omega_{L}^\prime
t,\,\mu\right)\,,
\\ \label{eq:yvelocity}
\upsilon_y(t)&=& -c\eta \sqrt{1-\varepsilon^2}\,
\mathrm{sn}\left(\omega_{L}^\prime t, \,\mu\right)\,,\\
\label{eq:zvelocity}
    \upsilon_z(t)&=& c-c(1-\beta_z)\,
    \mathrm{dn}\left(\omega_{L}^\prime t,\,
    \mu\right)\,.
\end{eqnarray}
Periodic properties of the Jacobian function (see eqs.
(\ref{eq:perprop}) in the  Appendix  \ref{sec:appendix2}) tell us
that the components of the charged particle's velocity in the plane
orthogonal to the wave propagation are periodic functions of time
with period
\begin{equation}\label{eq:partperod}
T_{P}:=\frac{4\mathrm{K}}{\omega_L^\prime}=\frac{2\pi}
{\omega_P}\,,
\end{equation}
while in the direction of propagation the oscillation's period is
half the size $T_\mathrm{P}/2\,.$ The fundamental circular
frequency of the particle's motion, $\omega_{\mathrm{\,P}}$,
differs from the frequency of the laser field
\begin{equation}\label{eq:new}
\omega_{P}=\frac{\pi}{2\mathrm{K}}\,\omega_{L}^\prime\,.
\end{equation}
From this expression we see that, apart from the pure kinematical
non-relativistic Doppler shift (\ref{eq:Doppl}), the frequency of
a particle's oscillation depends on the laser's intensity through
the real quarter period $\mathrm{K}$ (\ref{eq:KkK'}) of the
Jacobian functions. The presence of $\mathrm{K}$ in (\ref{eq:new})
exposes a new property of a particle's dynamics which is beyond
the dipole approximation. A particle oscillates at frequency that
depends on the laser intensity and polarization  in a nonlinear
way. For the low intensity regime, $\eta \ll 1\,,$ the period of a
particle's oscillation can be represent in the form of an
expansion (\ref{eq:Aperiodexp}) as
\begin{equation}\label{eq:smallfreq}
T_P=\frac{2\pi}{\omega_L^\prime}\left[
1+\left(\frac{1}{2}\right)^2\frac{1-2\,\epsilon^2}{(1-\beta_z)^2}\,\eta^2
+ \left(\frac{1\cdot3}{2\cdot4}\right)^2
\frac{(1-2\,\epsilon^2)^2}{(1-\beta_z)^4}\,\eta^4 + \dots
\right]\,.
\end{equation}

It is well-known that, in contrast to the dipole approximation
where the particle's motion is a pure harmonic, the relativistic dynamics
exhibits a drift action of the laser field on the particle which depends
on the intensity \cite{Sengupta1949,McMillan,Kibble1965}. In our
consideration, when the relativistic effects are partially taken
into account, this effect also can be seen.  For the
non-relativistic case with our choice of the initial conditions,
$\boldsymbol{\Pi}_\bot=0\,,$ the mean velocity for the transverse
direction vanishes
\begin{equation}\label{eq:trandrift}
\langle \boldsymbol{\upsilon}_\bot \rangle = 0\,.
\end{equation}
While the drift in the direction of propagation is a nonlinear
function of the laser beam's intensity
\begin{equation}\label{eq:meanvelocity}
\langle {\upsilon_{z}} \rangle =c - \frac{\pi
c}{2\mathrm{K}}(1-\beta_z)\,.
\end{equation}
This drift  velocity $\langle {\upsilon_{z}} \rangle$ for small
intensities at leading order reads
\begin{equation}\label{eq:meanvelocitysmall}
\langle {\upsilon_{z}} \rangle = \upsilon_z(0)\left(1-
\frac{1-2\,\epsilon^2}{4(1-\beta_z)^2}\,\eta^2 \right)+ \dots\,.
\end{equation}

Another new feature when comparing to the the dipole-approximation is
the appearance of the higher harmonics in the particle's motion. This
can be seen from our solution with the aid of the well-known Fourier
series expansion for the Jacobian function
\cite{WhittakerWatson,Bateman}. Using these  formulae collected in
Appendix  \ref{sec:appendix2} one can represent the trajectory as
\begin{eqnarray}\label{eq:harmonicstrajx}
  x(t) &=&\frac{4c\epsilon}{\omega_L\sqrt{1-2\,
\epsilon^2}}\,\sum_{n=1}^{\infty}\,\frac{q^{n-1/2}}{(2n-1)(1+q^{2n-1})}\,
  \sin(2n-1)
  \omega_{P}\,t\,,\\
  y(t)&=&\frac{8c\sqrt{1-\epsilon^2}}{\omega_L}\,
  \sum_{n=1}^{\infty}\,\frac{q^{n-1/2}}{(2n-1)(1-q^{2n-1})}\,
  \sin^2\left(n-\frac{1}{2}\right)
  \omega_{P}\,t \,, \\ \label{eq:harmonicstrajy}
  z(t) &=& \langle {\upsilon_{z}} \rangle\,t  -\frac{c}{\omega_L}\,\sum_{n=1}^{\infty}\,
  \frac{2q^n}{n(1+q^{2n})}\,\sin 2n\,\omega_{P}\,t\,,\label{eq:harmonicstrajz}
\end{eqnarray}
where $q$ is the so-called {\em nome} parameter
\begin{equation}
q:=\exp\left(-\pi\,\frac{\mathrm{K}^\prime}{\mathrm{K}}\right)\,.
\end{equation}
Note (see eq. (\ref{eq:Anome})) that the nome $q$
for small intensities is approximately
\begin{equation}\label{eq:1appr}
    q\approx \frac{1-2\,\epsilon^2}{16(1-\beta_z)^2}\,\eta^2 +
    O(\eta^4)\,.
\end{equation}

When the intensity parameter is small one can perform a Galilean
boost with the velocity $\boldsymbol{V}:=-\left(0, 0, \langle
{\upsilon_{z}} \rangle \right)\,$ to the so-called \textit{average
rest frame} (ARF), frame where the mean particle's velocity
vanishes. In this frame the particle's motion represents only the
superposition of harmonic oscillations with the fundamental
frequency $\omega_{P}\,.$

In the ARF frame equations
(\ref{eq:harmonicstrajx})-(\ref{eq:harmonicstrajz}) reduce, for
small intensities, to the following expressions in the leading
order of the $\eta$-expansion
\begin{eqnarray}\label{eq:smaletax}
x_{{}_\mathrm{ARF}}(t)&=&-\frac{c\varepsilon
}{\omega_L^\prime}\,\eta \sin\omega_L^\prime t \,,
\\
\label{eq:smaletay}
y_{{}_\mathrm{ARF}}(t)&=-&\frac{c\sqrt{1-\varepsilon^2}}{\omega_L^\prime}\,\eta
\left(1-\cos\omega_L^\prime t \right),
\\
\label{eq:smaletaz}
z_{{}_\mathrm{ARF}}(t)&=&\frac{c}{\omega'_L}\,\frac{1-2\,\epsilon^2}{8(1-\beta_z)}\,\eta^2
\,\sin 2\omega_L^\prime t\,.
\end{eqnarray}
Pictorially  these formulae describe the orbits  shaped like  a
figure of eight.  This form of the orbit is well-known from
the parametric relativistic solution
\cite{LandauLifshitz,ItzyksonZuber,Thirring} and
(\ref{eq:smaletax}) -(\ref{eq:smaletaz}) are the small intensity
approximation to the explicit representation of the  relativistic
trajectories. We omit here the detailed comparison with the
relativistic solution since it  requires the knowledge of a
relativistic particle's trajectory as a function of the physical
time as well as a careful analysis of the frame dependence of our
partially relativistic solution. We intend to do this  in a
forthcoming publication.

All mentioned features, the Doppler shift, dependence of the
particle's oscillation frequency  on the laser beam's intensity as
well as the presence of higher harmonics in the particle's motion, lead
to a several important phenomena. Among them there are the non-linear
modification to the classical Thomson scattering and a charged
particle's mass/energy shift in the electromagnetic background
radiation.

%%%%%%%%%%%%%%%%%%%%%%%%%%%%%% V
\section{Concluding remarks}
%%%%%%%%%%%%%%%%%%%%%%%%%%%%%%%%%%

Most textbooks on the classical dynamics of particles pay tribute to a
historical tradition. They begin with the consideration of
non-relativistic mechanics and, after introducing the basics of relativity,
discuss the specific features of the corresponding
relativistic problem. The exception to this rule is the charged
particle's dynamics in the background of an electromagnetic plane wave. All
well-known sources either discuss the problem in the dipole
approximation \cite{Heitler} or solve the problem in the framework
of relativistic mechanics directly
\cite{LandauLifshitz,ItzyksonZuber,Thirring}.

In the present note we ``restore historical  justice'' by
considering the classical problem of a Newtonian particle's motion
in a given electromagnetic  plane wave field  beyond the above
mentioned dipole approximation. We obtain an exact representation
for a particle's orbit in the parametric form which is analogous
to the well-known relativistic solution. Furthermore, we show that
the three-dimensional non-relativistic orbit of a particle's
motion in a monochromatic arbitrarily polarized plane wave admits,
as an explicit function of  the laboratory frame's time, the exact
solution  in terms of the doubly periodic elliptic functions. This
is in contrast to the relativistic trajectory which is known
explicitly only in a four dimensional parametric form. The derived
solution explicitly exposes the presence of higher fundamental
harmonics in a charged particle's motion as well as nonlinear
dependence of the oscillations on the intensity and polarization
of the monochromatic background. Another interesting
characteristic  of the trajectory we discovered is an intensity
duality between motion in different radiation backgrounds. Since a
given intensity defines the modulus  of the Jacobian functions,  a
particle's trajectory in backgrounds whose intensities are
connected by the modular transformations are simply related.

Finally it is worth commenting on the sinuous way to derive the
exact particle trajectories  within the  reparametrization
invariant form of the Hamilton-Jacobi method presented here. We
prefer to follow it since  this approach incorporates
both a classical as well as a quantum treatment. The knowledge of
the classical generating function connecting the interacting and free
systems helps to construct the corresponding quasiclassical
description and in particular can be exploited for  study of the
laser-atom interactions \cite{books}. Furthermore, the suggested
approach mimics the four dimensional relativistic
treatment which is useful to study the deviation from
non-relativistic motion. All these issues  we intend to discuss in
detail in future  publications. Apart from this, based on the
explicit form of the non-relativistic trajectory we plan to
study several important effects, including the non-linear
Thomson/Compton scattering, electromagnetic dressing, charged
particle acceleration by a laser field etc., in the transition
regimes between the non-relativistic and relativistic cases.

%%%%%%%%%%%%%%%%%%%%%%%%%%%%    Acknowledgments
\begin{acknowledgments}
%%%%%%%%%%%%%%%%%%%%%%%%%%%%%%%%%%%%%%%%%%%%%%%%

We are grateful to M.~Eliashvili, V.~Gerdt, T.~Heinzl,  P.~Jermey,
M.~Lavelle, D.~McMullan, and S.~Vinitsky for a productive discussions.
This work was supported in part by the Ministry of
Education and Science of the Russian Federation, grant \#
1027.2008.2 and by the Georgian National Science Foundation, grant
\# GNSF/ST06/4-050.
\end{acknowledgments}

%%%%%%%%%%%%%%%%%%%%%%% APPENDICES
\appendix
%%%%%%%%%%%%%%%%%%%%%%%%%%%%%%%%%%%%%%%%%%%%

%%%%%%%%%%%%%%%%%%%%%%%%%%%%%%%%%%%%%%%%%%%%%%%%%%%%%%%%%%%%%%%%%%%%%%%%%%
\section{Remarks on ``non-relativistic'' symmetries } \label{sec:appendix1}
%%%%%%%%%%%%%%%%%%%%%%%%%%%%%%%%%%%%%%%%%%%%%%%%%%%%%%%%%%%%%%%%%%%%%%%%%%%

Below we discuss the realization of the gauge symmetry as well as
the implementation of the Galilean boost  for a
 charged  ``non-relativistic''  particle travelling in an
electromagnetic background.

%-------------------------------------- A.1
\subsection{Gauge transformations }
%------------------------------------------

In terms of the  3-dimensional notations for a gauge potential
\(A^\mu=(\Phi\,,\, \boldsymbol{A}) \) and coordinates
$x^\mu=(ct\,,x,y,z)= (ct\,,\boldsymbol{x})$, the gauge
transformation reads
\begin{eqnarray}\label{eq:gtransf}
\boldsymbol{A}(t, \boldsymbol{x}) \to \boldsymbol{A}^\prime(t,
\boldsymbol{x}) &=& \boldsymbol{A}(t, \boldsymbol{x}) +
\frac{\partial}{\partial \boldsymbol{x} }\, \Omega(t,
\boldsymbol{x})
\,,\\
\Phi(t, \boldsymbol{x}) \to \Phi^\prime(t, \boldsymbol{x})&=&
\Phi(t, \boldsymbol{x}) -\frac{1}{c}\frac{\partial}{\partial t}\,
\Omega(t, \boldsymbol{x}) \,.\label{eq:gtransf2}
\end{eqnarray}
Under the  transformations (\ref{eq:gtransf}) and
(\ref{eq:gtransf2}) the ``non-relativistic''
 Lagrangian
(\ref{eq:nrL}) \textit{is not invariant}:
\begin{equation}\label{eq:nrLchange}
    \mathcal{L}\left(\boldsymbol{x}\,,
    \frac{\mathrm{d}\boldsymbol{x}}{\mathrm{d}t}\,, t\right) \to
\mathcal{L}\left(\boldsymbol{x}\,,
    \frac{\mathrm{d}\boldsymbol{x}}{\mathrm{d}t}\,, t\right) +
    \frac{e}{c}\,\frac{\mathrm{d}}{\mathrm{d}
t}\Omega(t\,, \boldsymbol{x}(t))\,.
\end{equation}
Here  ${\mathrm{d}}/{\mathrm{d} t}$ denotes the total derivative
\begin{equation}
    \frac{\mathrm{d}}{\mathrm{d}
t}:=\frac{\mathrm{d}\boldsymbol{x}}{\mathrm{d}t}\cdot\,\frac{\partial}{\partial
\boldsymbol{x} } +\frac{\partial}{\partial t}\,.
\end{equation}
However, the variation (\ref{eq:nrLchange}) being the total
derivative of the function \-${e}/{c}\,\Omega( t\,,
\boldsymbol{x}(t) )\,,$ {does not affect} to the classical
equations of motion.

The changes of  gauge potentials (\ref{eq:gtransf}) and
(\ref{eq:gtransf2}) are canonical transformations with the
generating function
\begin{equation}\label{eq:gencantr}
    F(\boldsymbol{x},\boldsymbol{P}, t) = \boldsymbol{x}\cdot\boldsymbol{P}-\Omega( t\,,
\boldsymbol{x}(t) )\,.
\end{equation}

%---------------------------------
\subsection{Galilean boosts} \label{ap:Galboosts}
%------------------------------------

Under a Lorentz boost in the $x$ direction with the factor $\gamma:
=1/\sqrt{1-{V^2}/{c^2}}$
\begin{equation}\label{eq:cLboost}
  t^\prime=\gamma\left(t-\displaystyle\frac{V}{c^2}\,x\right)\,, \qquad
  x^\prime= \gamma\left(x-Vt\right)\,,\qquad
  y^\prime=y\,,
  \qquad z^\prime=z\,,
\end{equation}
a gauge potential $A^\mu=(\Phi, \boldsymbol{A}) $ being a Lorentz
4-vector  transforms as
\begin{equation}
 \Phi^\prime=\gamma\left(
    \Phi-\displaystyle\frac{V}{c}\,A_x\right)\,,\qquad
    {A_x}^\prime= \gamma\left(A_x-\displaystyle\frac{V}{c}\,\Phi\right)\,,
\qquad {A_y}^\prime={A_y}\,,\qquad {A_z}^\prime={A_z}\,.
\end{equation}
For small velocities, neglecting terms of order $V^2/c^2$ and
higher the Lorentz boost (\ref{eq:cLboost}) reduces to the
Galilean boost
\begin{equation}
     t^\prime= t \,, \qquad x^\prime={x-Vt}\,,\qquad y^\prime=y\,,\quad
z^\prime=z\,,
\end{equation}
while the  gauge potential changes as
\begin{equation}
 \Phi^\prime={\Phi-\frac{V}{c}\,A_x}\,,\qquad
    {A_x}^\prime={A_x-\frac{V}{c}\,\Phi}\,,
\qquad {A_y}^\prime={A_y}\,,\qquad {A_z}^\prime={A_z}\,.
\end{equation}

With the help of these relations, keeping only the leading terms in $V/c$, we have
\begin{eqnarray}\label{ChLB}
\mathcal{L}
  \to \mathcal{L}^\prime &=&\mathcal{L}-m\frac{\mathrm{d}}{\mathrm{d}t}
  \left(V\,x -\frac{1}{2}\,V^2t \right)\,,
\end{eqnarray}
note that $V$ is \textit{not} infinitesimally
small.
The Lagrangian is invariant under the  Galilean boost {\em up to a
total derivative } only.

Both variations of the Lagrangian,  under the gauge
transformations (\ref{eq:nrLchange}) as well as under the Galilean
boost (\ref{ChLB}), are not important at the classical level but can
lead in general to a nontrivial quantum phenomena. Particularly,
they lead to the appearance of the so-called 1-cocycle for the
wave function. The action of the unitary operator $U_G$ generating
the Galilean boost on the wave function reads
\begin{equation}\label{eq:qu1coc}
    U_G\Psi(\boldsymbol{x})=\exp{i\left(m\boldsymbol{V}\cdot\boldsymbol{x}
    -\frac{1}{2} m \boldsymbol{V}^2t
    \right)}\,\Psi(\boldsymbol{x}-\boldsymbol{V}t)\,.
\end{equation}
It is interesting that for the charged particle  this cocycle can
be trivialized. Indeed, if the Galilean boost is accompanied by
the gauge transformation generated by (\ref{eq:gencantr}) with the
special gauge function
$$\Omega(t, \boldsymbol{x}(t)):= \frac{e}{c}\,\left(m\boldsymbol{V}\cdot\boldsymbol{x}
    -\frac{1}{2} m \boldsymbol{V}^2t \right)\,,
$$
then  the Lagrangian (\ref{eq:nrL}) remains unchanged and cocycle
is removed.

%%%%%%%%%%%%%%%%%%%%%%%%%%%%%%%%%%%%%%%%%%%%%%%%%%%%%%%%%%%%%%
\section{Generic boundary conditions} \label{ap:genericbound}
%%%%%%%%%%%%%%%%%%%%%%%%%%%%%%%%%%%%%%%%%%%%%%%%%%%%%%%%%%%%%%%%%%

Here we briefly state  how one can express the physical time $t$
in terms of the auxiliary evolution parameter when  the generic
boundary conditions on a particle's position and velocity are imposed.
To find this dependance one can proceed as follows. Plugging the
expression for the gauge potential (\ref{eq:gpot}) into the function
$W(u, \Pi_\bot)\, $ from (\ref{eq:w}) we have
\begin{equation*}\label{eq:wmono}
W(u, \Pi_\bot)=-\eta^2m^2c^2\left(\epsilon^2\cos^2
u+(1-\epsilon^2)\sin^2 u\right) +2\eta mc \left(\Pi_1\cos u
+\Pi_2\sin u\right)\,.
\end{equation*}
The integral in (\ref{eq:1nparsolution}) with this expression
represents the so-called elliptical integral. Indeed, using the
universal trigonometric substitution $x=\tan(u/2)$ it can be
rewritten in the Weierstrass form:
\begin{equation}\label{eq:timegen}
 t(s) = \frac{2}{\omega_L}\,\int_0^{\tan(\omega_Ls/2)} \mathrm{d}x\,
 \frac{1}{\sqrt{f(x)}}\,,
\end{equation}
with the fourth order polynomial
\begin{equation}\label{eq:4pol}
    f(x)= a_0x^4+4a_1x^3+6a_2x^2+4a_3x+a_4\,,
\end{equation}
whose  coefficients are
\begin{eqnarray}\label{eq:coeffipol}
&&a_0:=\beta_+^{2} - \eta ^{2}\,\varepsilon^{2} -
2\,\eta\,\beta_1\,,\qquad a_{1}:= \eta\,\beta_2\,,\qquad a_{3}:= \eta\,\beta_2\,, \\
&& a_2 := \frac{1}{3}\beta_+^{2}+ \eta^{2}\,\left(\varepsilon ^{2}
-\frac{2}{3}\right), \qquad a_4:=\beta_+^{2} - \eta
^{2}\,\varepsilon^{2}+ 2\,\eta\,\beta_1\,.
\end{eqnarray}
According to the classical result (see e.g.
\cite{WhittakerWatson}) attributed to Weierstrass, the integral
(\ref{eq:timegen}) can be inverted:
\begin{equation}\label{eq:invert}
 \tan\left(\frac{\omega_Ls}{2}\right)   =
\frac{\displaystyle{\sqrt{f(0)}\,\wp^\prime\left(\frac{\omega_L}{2}\,t\right)-\frac{1}{2}
\left[\wp\left(\frac{\omega_L}{2}\,t\right)-\frac{1}{24}f^{''}(0)\right]+
\frac{1}{24}f(0)f^{'''}(0)}}
{\displaystyle{2\left[\wp\left(\frac{\omega_L}{2}\,t\right)-\frac{1}{24}f^{''}(0)\right]^2
-\frac{1}{48}f(0)f^{''''}(0)}}\,.
\end{equation}
Here the number of primes over the polynomial (\ref{eq:4pol})
denotes the order of the derivatives with respect to $x$. In
(\ref{eq:invert}) the Weierstrass doubly periodic function
$\wp\left(z\,; g_2\,, g_3 \right)$ depends on two invariants $g_2$
and $g_3$ of the polynomial (\ref{eq:4pol})
\begin{eqnarray*}\label{eq:polinvar}
    g_2:&=&4\,\eta^4\left(\varepsilon^4-\varepsilon^2+\frac{1}{3}\right)
-4\eta^2\left(\frac{1}{3}\beta_+^2+
\boldsymbol{\beta}^2_{\bot}\right)+\frac{4}{3}\beta_+^4\,,\\
g_3:&=&
\frac{4}{3}\eta^6\left(\varepsilon^4-\varepsilon^2+\frac{2}{9}\right)
- 4\eta^4\left[\frac{2}{3}\,\varepsilon^4\beta_+^2 +
\varepsilon^2\left(\beta_1^2-\beta_2^2-\frac{2}{3}\beta_+^2\right)
+\frac{1}{9}\beta_+^2
-\frac{2}{3}\,\beta_1^2+\frac{1}{3}\,\beta_2^2
\right]\\
&-&\frac{4}{3}\,\eta^2\beta_+^2\left(\frac{1}{3}\beta_+^2+\boldsymbol{\beta}^2_\bot\right)+
\frac{8}{27}\,\beta_+^6\,.
\end{eqnarray*}
These equations show that the dependence of any particle's
trajectory on the initial conditions is accumulated in the
invariants $g_2, g_3$ and functionally is quite subtle.
Particulary there is no rotational degeneracy with respect to the
vector $\boldsymbol{\Pi_{\bot}}\,.$

%%%%%%%%%%%%%%%%%%%%%%%%%%%%%%%%%%%%%%%%%%%%%%%%%%%%%%%%%%%%%%%%%%%%%%%%%%%%%
\section{Vocabulary on the Jacobian elliptic functions}\label{sec:appendix2}
%%%%%%%%%%%%%%%%%%%%%%%%%%%%%%%%%%%%%%%%%%%%%%%%%%%%%%%%%%%%

Following the classical textbooks \cite{WhittakerWatson,Bateman}
and handbook \cite{AbramowitzStegun} we collect the basic formulae
on Jacobian elliptic functions which are extensively used in the
main text.

The \textit{amplitude function} $\phi:=\mathrm{am}(z\,, \mu)$ is
the \textit{inverse } of the function  defined  by the integral
\begin{equation}\label{eq:amplitude}
z(\phi) =\int_0^\phi\,\mathrm{d}
\vartheta\frac{1}{\sqrt{1-\mu^2\sin^2\vartheta}}\,.
\end{equation}
The amplitude $\mathrm{am}(z\,, \mu)$ is an infinitely many-valued
function whose principal domain of definition for real $z$ is
$(-\mathrm{K}\,, \mathrm{K})\,.$ The value of this constant
$\mathrm{K}$ is determined by the  \textit{modulus} $\mu$ via the
so-called \textit{ complete elliptic integral}, see
(\ref{eq:KkK'}) below.

Three basic  \textit{Jacobian elliptic functions},
$\mathrm{sn}(z,\mu),\, \mathrm{cn}(z,\mu) $ and
$\mathrm{dn}(z,\mu)$ are analytical functions of the complex variable
$z$ everywhere except at the simple poles. These elliptic function are
expressible in terms of the amplitude function
\begin{equation}\label{jacobi:def}
\mathrm{sn}(z, \mu)=\sin\big(\mathrm{am}(z\,, \mu)\big)\,, \quad
\mathrm{cn}(z, \mu)=\cos\big(\mathrm{am}(z\,, \mu)\big)\,, \quad
\frac{\mathrm{d}}{\mathrm{d}z}\,\mathrm{am}(z,\mu)=
\mathrm{dn}(z,\mu)\,,
\end{equation}
and satisfy the basic  algebraic
\begin{equation}\label{eq:basicrel}
    \mathrm{sn}^2z+\mathrm{cn}^2z=1\,,\quad \mu^2\mathrm{sn}^2z+\mathrm{dn}^2z=1\,,
\end{equation}
and differential relations
\begin{eqnarray} \label{jacobi:derivatives}
\frac{\mathrm{d}}{\mathrm{d}z}\,\mathrm{sn}(z,\mu) &=&
\mathrm{cn}(z\,,\mu)\,\mathrm{dn}(z\,,\mu)
\,,\\
\frac{\mathrm{d}}{\mathrm{d}z}\,\mathrm{cn}(z,\mu)
&=&-\mathrm{sn}(z\,,\mu)\,\mathrm{dn}(z\,,\mu)
\,,\\ \label{jacobi:derivatives2}
\frac{\mathrm{d}}{\mathrm{d}z}\,\mathrm{dn}(z,\mu) &=&-\mu^2
\mathrm{sn}(z\,,\mu)\,\mathrm{cn}(z\,,\mu) \,,
\end{eqnarray}
which show the analogy of the Jacobian functions to the
trigonometric functions.

Functions $\mathrm{sn}(z,\mu),\, \mathrm{cn}(z,\mu) $ and
$\mathrm{dn}(z,\mu)$ are doubly periodic functions of $z$.
Periods of $\mathrm{sn}(z,\mu)$ are $4\mathrm{K}$ and
$2i\mathrm{K}^\prime\,$ while periods of $\mathrm{cn}(z,\mu)$ are
$4\mathrm{K}$ and $2\mathrm{K}+2i\mathrm{K}^\prime\,.$ Function
$\mathrm{dn}(z,\mu)$ has periods $2\mathrm{K}$ and
$4i\mathrm{K}^\prime\,.$  The ``real'' $(\mathrm{K})\,$ and
``imaginary'' $(\mathrm{K}^\prime)\,$ quarter periods are real
numbers given by the complete elliptic integrals
\begin{eqnarray}\label{eq:KkK'}
\mathrm{K}(\mu)&:=&\int_0^{\pi/2}\,\mathrm{d}\vartheta
\frac{1}{\sqrt{1-\mu^2\sin^2\vartheta}}\,, \\
i\,\mathrm{K}^\prime(\mu)&:=&i\,\int_0^{\pi/2}\,\mathrm{d}
\vartheta
\frac{1}{\sqrt{1-(1-\mu^2)\sin^2\vartheta}}\,.\label{eq:KkK'2}
\end{eqnarray}
The Jacobian functions as functions of the modulus  are single
valued on the complex $\mu $ plane  with two cuts $[1, \infty )\,$
and $(-\infty, 0]\,.$

\underline{\textit{The discrete symmetry.}} The Jacobian functions
$\mathrm{cn}$ and $\mathrm{dn}$ are even functions , while
$\mathrm{sn}$ is odd. They obey the relations
\begin{eqnarray}\label{eq:perprop}
\mathrm{sn}(u +2m\mathrm{K}+2ni\mathrm{K}^\prime\,, \mu )&=& (-)^m\,\mathrm{sn}(u\,, \mu )\,, \\
\mathrm{cn}(u
+2m\mathrm{K}+2ni\mathrm{K}^\prime\,,\mu)&=&(-)^{m+n}\,\mathrm{cn}(u\,,
\mu)\,,\\
 \mathrm{dn}(u +2m\mathrm{K}+2ni\mathrm{K}^\prime\,, \mu )&=& (-)^n\,\mathrm{dn}(u\,, \mu )\,,
\end{eqnarray}
where $n, m \in \mathbb{Z}\,.$

\underline{\textit{Two degenerate cases.}} The doubly periodic
Jacobian functions degenerate to other functions when one of the periods become
infinite, that if $\mu$  is $0$ or $1\,.$

When $\mu= 1\, $ the real quarter-period  $\mathrm{K}=\infty $ and
Jacobian function degenerate to the hyperbolic functions
\begin{eqnarray}\label{eq:sndegmu1}
  \mathrm{sn}(u, 1) = \tanh u\,, \qquad
   \mathrm{cn}(u, 1)=\frac{1}{\cosh u} \,, \qquad
 \mathrm{dn}(u, 1)=\frac{1}{\cosh u}\,.
\end{eqnarray}

If   $\mu= 0\, $ the real quarter-period  is finite,
$\mathrm{K}=\pi/2\,,$ but the imaginary quarter-period  is
infinite and the Jacobian function degenerate to the trigonometric
functions
\begin{eqnarray}\label{eq:sndegmu2}
  \mathrm{sn}(u, 0) = \sin u\,, \qquad
   \mathrm{cn}(u, 0)=\cos u \,, \qquad
 \mathrm{dn}(u, 0)= 1 \,,
\end{eqnarray}

\underline{\textit{Small modulus expansions.}} When modulus is
small enough the Jacobian functions can be approximated as
\begin{eqnarray}
\mathrm{am}(z,\mu)&=& z -
\frac{1}{4}\,\mu^2\left[z-\sin(z)\cos(z)\right] + o(\mu^4)\,,\\
\mathrm{sn}(z,\mu)&=&\sin(z) -
\frac{1}{4}\,\mu^2\left[z-\sin(z)\cos(z)\right]\cos(z) +
o(\mu^4)\,,\\
 \mathrm{cn}(z,\mu)&=&\cos(z) +
\frac{1}{4}\,\mu^2\left[z-\sin(z)\cos(z)\right]\cos(z) +
o(\mu^4)\,,\\
\mathrm{dn}(z,\mu)&=& 1-\frac{1}{2}\,\mu^2\sin^2(z) + o(\mu^4)\,.
\end{eqnarray}
The quarter period $\mathrm{K}$ has a small modulus expansion of
the form
\begin{equation}\label{eq:Aperiodexp}
\mathrm{K} =\frac{\pi}{2}\left[ 1+\left(\frac{1}{2}\right)^2\mu^2
+ \left(\frac{1\cdot3}{2\cdot4}\right)^2\mu^4 +
\left(\frac{1\cdot3\cdot5}{2\cdot4\cdot6}\right)^2\mu^6+ \dots
\right]\,.
\end{equation}

\underline{\textit{The Fourier series expansions:}} These are
\begin{eqnarray}\label{eq:FLambertSer}
  \mathrm{sn}(z,\, \mu)&=& \frac{2\pi}{\mu\mathrm{K}}\,
  \sum_{n=1}^{\infty}\,\frac{q^{n-1/2}}{1-q^{2n-1}}\,
  \sin(2n-1)\frac{\pi z}{2\mathrm{K}}\,,\\
 \mathrm{cn}(z,\, \mu) &=&\frac{2\pi}{\mu\mathrm{K}}\,
  \sum_{n=1}^{\infty}\,\frac{q^{n-1/2}}{1+q^{2n-1}}\,
  \cos(2n-1)\frac{\pi z}{2\mathrm{K}}\,, \\
   \mathrm{dn}(z,\, \mu) &=& \frac{2\pi}{\mathrm{K}}+
   \frac{2\pi}{\mathrm{K}}\,\sum_{n=1}^{\infty}\,
  \frac{q^n}{1+q^{2n}}\,\cos 2n\,\frac{\pi z}{2\mathrm{K}}\,,
\end{eqnarray}
with the so-called {\em nome}, or Jacobi parameter
$q=\exp\left(-\pi{\mathrm{K}^\prime}/{\mathrm{K}}\right)\,.$
Similarly to the quarter periods the nome $q$ can be expanded in
powers of the modulus
\begin{equation}\label{eq:Anome}
q=\frac{\mu^2}{16} + 8\,\left(\frac{\mu^2}{16}\right)^2
+84\,\left(\frac{\mu^2}{16}\right)^3+ \dots\,,
\end{equation}

\underline{\textit{The modular transformations.~}} The ratio $\tau
:=i{\mathrm{K}^\prime}/{\mathrm{K}}$ serves as an important
modular parameter\footnote{We assume that $\mathrm{Re}{(\tau)}>
0\,.$}.
 Under the  unimodular transformation
\begin{equation}\label{eq:modulartr}
    \tau \to \frac{c+d\tau}{a+b\tau}\,, \qquad a,b,c,d \in
    \mathbb{Z}\,, \qquad ad-bc=1\,,
\end{equation}
the doubly periodic elliptic functions with different periods are
expressible through each other.

Any transformation (\ref{eq:modulartr}) can be represented as a
product of powers of two generators
\begin{equation}\label{eq:genunimodul}
\mathrm{S}:= \left[%
\begin{array}{cc}
  0 & 1 \\
  -1 & 0 \\
\end{array}%
\right]\,, \qquad  \mathrm{T}  := \left[%
\begin{array}{cc}
  1 & 0 \\
  1 & 1 \\
\end{array}%
\right] \,.
\end{equation}
The matrix $\mathrm{S}$, generates  the so-called Jacobi imaginary
transformation
$$\tau \to \tau^\prime: = -1/\tau\,,
$$
under which the Jacobian functions vary as
\begin{eqnarray}\label{eq:Jacobitran1}
\mathrm{sn}(iz, \mu^\prime)=i\frac{\mathrm{sn}(z,
\mu)}{\mathrm{cn}(z, \mu)}, \quad \mathrm{cn}(iz,
\mu^\prime)=\frac{1}{\mathrm{cn}(z, \mu)},\quad \mathrm{dn}(iz,
\mu^\prime)=\frac{\mathrm{dn}(z, \mu)}{\mathrm{cn}(z, \mu)},\qquad
\end{eqnarray}
where $\mu^\prime:=\sqrt{1-\mu^2}$, the \textit{complementary
modulus}.
The action of the generator  $\mathrm{T}$ which represents the shift
transformation
\[
\tau \to \tau^\prime: =1+\tau\,,
\]
results in the following relations between the basic Jacobian
functions
\begin{eqnarray}\label{eq:shifttran21}
\mathrm{sn}(\mu^\prime z, \frac{i\mu}{\mu^\prime})= \mu^\prime
\frac{\mathrm{sn}(z, \mu)}{\mathrm{dn}(z, \mu)}, \quad
\mathrm{cn}(\mu^\prime z\,, \frac{i\mu}{\mu^\prime})=
\frac{\mathrm{cn}(z, \mu)}{\mathrm{dn}(z, \mu)}, \quad
\mathrm{dn}(\mu^\prime z, \frac{i\mu}{\mu^\prime})=
\frac{1}{\mathrm{dn}(z, \mu)}. \qquad
\end{eqnarray}
Finally the change
\[\tau \to \tau^\prime:=\tau/{(1+\tau)}\,,
\]
which can be represented as,  $\mathrm{TST}$, is of special
interest to us since it relates the functions with the inverse moduli
$\mu$ and $\underline{\mu}:={\mu}^{-1}$:
\begin{eqnarray}\label{eq:JacobitranC}
\mathrm{sn}(\mu z\,,\underline{\mu} )= \mu\, \mathrm{sn}(z\,,
\mu)\,, \quad \mathrm{cn}(\mu z\,, \underline{\mu})=
\mathrm{dn}(z\,, \mu)\,, \quad  \mathrm{dn}(\mu z\,,
\underline{\mu})= \mathrm{cn}(z\,, \mu)\,.\quad
\end{eqnarray}

\newpage
%%%%%%%%%%%%%%%%%%%%%%%%%%%%%%%%%%%%% REFERENCES %%%%%%%%%%%%%%%%

\end{document}